\documentclass[aps,a4paper,twocolumn,english,superscriptaddress,eqsecnum,longbibliography]{revtex4-1}

\usepackage{enumitem}
\usepackage{graphicx}
\usepackage{amssymb}
\usepackage{amsmath}
\usepackage{mathtools}
\usepackage{xcolor}

\usepackage[colorlinks=true,
            linkcolor=red,
            citecolor=blue,
            urlcolor=blue]{hyperref}

\begin{document}

\title{From eigenstate to Hamiltonian: Prospects for ergodicity and localization}

\author{Maxime Dupont}
\affiliation{Department of Physics, University of California, Berkeley, California 94720, USA}
\affiliation{Materials Sciences Division, Lawrence Berkeley National Laboratory, Berkeley, California 94720, USA}

\author{Nicolas Mac\'e}
\affiliation{Laboratoire de Physique Th\'eorique, IRSAMC, Universit\'e de Toulouse, CNRS, UPS, 31062 Toulouse, France}

\author{Nicolas Laflorencie}
\affiliation{Laboratoire de Physique Th\'eorique, IRSAMC, Universit\'e de Toulouse, CNRS, UPS, 31062 Toulouse, France}

\begin{abstract}
    This paper addresses the so-called inverse problem which consists in searching for (possibly multiple) parent target Hamiltonian(s), given a single quantum state as input. Starting from $\Psi_0$, an eigenstate of a given local Hamiltonian $\mathcal{H}_0$, we ask whether or not there exists another parent Hamiltonian $\mathcal{H}_\mathrm{P}$ for $\Psi_0$, with the same local form as $\mathcal{H}_0$. Focusing on one-dimensional quantum disordered systems, we extend the recent results obtained for Bose-glass ground states [M. Dupont and N. Laflorencie, \href{https://journals.aps.org/prb/abstract/10.1103/PhysRevB.99.020202}{Phys. Rev. B \textbf{99}, 020202(R) (2019)}] to Anderson localization, and the many-body localization (MBL) physics occurring at high energy. We generically find that any localized eigenstate is a very good approximation for an eigenstate of a distinct parent Hamiltonian, with an energy variance $\sigma_\mathrm{P}^2(L)=\langle\mathcal{H}_\mathrm{P}^2\rangle_{\Psi_0}-\langle\mathcal{H}_\mathrm{P}\rangle_{\Psi_0}^2$ vanishing as a power law of system size $L$. This decay is microscopically related to a chain-breaking mechanism, also signalled by bottlenecks of vanishing entanglement entropy. A similar phenomenology is observed for both Anderson and MBL. In contrast, delocalized ergodic many-body eigenstates uniquely encode the Hamiltonian in the sense that $\sigma_\mathrm{P}^2(L)$ remains finite at the thermodynamic limit, i.e., $L\to+\infty$. As a direct consequence, the ergodic-MBL transition can be very well captured from the scaling of $\sigma_\mathrm{P}^2(L)$.
\end{abstract}

\maketitle

\section{Introduction}

Engineering trial wave functions to capture relevant quantum physical phenomena is often the first step in understanding them. It is also the basis of variational approaches where only a handful of free parameters need to be optimized or fitted in order to accurately reproduce observations. This approach has proven very successful in condensed matter, with the famous examples of the Bardeen-Cooper-Schrieffer theory of superconductivity~\cite{bardeen1957} and the Laughlin wave function used to explain the fractional quantum Hall effect~\cite{laughlin1983}, to cite but a few. The next step is to ensure that such an ansatz is indeed a good approximation of the actual ground state of the microscopic model describing the many-body system, or more generally to find a parent Hamiltonian for which the trial state would be the ground state.

Typically, given a quantum state $\Psi_0$, finding a parent Hamiltonian $\mathcal{H}_\mathrm{P}$ with reasonable physical origin and properties such as locality and pairwise interactions is a tremendous task. There is no guarantee of existence or uniqueness for that parent Hamiltonian, and one usually satisfies oneself with $\Psi_0$ being a \textit{decent} approximation of an eigenstate of $\mathcal{H}_\mathrm{P}$, if not the exact one. The quality of the approximation can be quantified in various ways: by the overlap of the trial state with the exact ground state of the parent Hamiltonian (which can be computed for small system sizes by means of exact diagonalization for instance) or by measuring the energy variance $\sigma^2_\mathrm{P}=\langle\mathcal{H}_\mathrm{P}^2\rangle_{\Psi_0}-\langle\mathcal{H}_\mathrm{P}\rangle_{\Psi_0}^2\geq 0$, equal to zero if $\Psi_0$ is an eigenstate of $\mathcal{H}_\mathrm{P}$, although not necessarily its ground state.

In this paper, we are interested in the properties of parent Hamiltonians for one-dimensional disordered systems. In particular, starting from an exact eigenstate $\Psi_0$ --- ground and excited states are considered --- of a disordered many-body Hamiltonian $\mathcal{H}_0$, characterized by its disorder configuration, we ask if there exists another parent Hamiltonian $\mathcal{H}_\mathrm{P}$ for $\Psi_0$, which would only differ from the original Hamiltonian by its disorder configuration. Because we start from an exact eigenstate, this task is related to the following question: Can a given eigenstate code for a unique local Hamiltonian? Recent studies have shown that the answer is generically positive~\cite{qi2017}, and holds in particular for eigenstates of disordered Hamiltonians, provided they satisfy the eigenstate thermalization hypothesis (ETH)~\cite{garrison2018}. Some of the authors of the current paper have provided strong numerical evidences that this statement no longer holds for localized many-body ground states~\cite{dupont2019}, for instance describing the Bose-glass state~\cite{Giamarchi1988,fisher_boson_1989}, even if one requires the parent Hamiltonian to keep the same local form.
Here, we extend this study by considering both ground and excited states for various one-dimensional disordered models, addressing both Anderson~\cite{RevModPhys.80.1355} and many-body localization (MBL) problems~\cite{nandkishore_many-body_2015,alet_many-body_2018,abanin_many-body_2019}.

We first investigate the random-field spin-half XY model in one dimension, which can be mapped to a disordered free-fermion model, and exhibits Anderson localization~\cite{mott1961} at all energy densities. Because of the free-fermion nature of the Hamiltonian, fairly large system sizes $L\sim 10^3$ can be accessed numerically. We then turn our attention to the interacting counterpart of the previous model, namely, the random-field spin-$1/2$ Heisenberg chain. The ground state of this model is always localized in the presence of disorder~\cite{Giamarchi1988,doty_effects_1992}, and this quantum phase of matter is known as Bose glass~\cite{Giamarchi1988,fisher_boson_1989}. However, at finite energy density there exists a mobility edge between a thermal and localized phase~\cite{kjall_many-body_2014,laumann_many_2014,luitz_many-body_2015,PhysRevB.92.064203}. The presence of interaction makes the numerical study much more challenging, and we are able to access via the exact shift-invert method~\cite{luitz_many-body_2015,pietracaprina2018} system sizes up to $L = 22$ spins. In both cases, we base our approach on the ``eigenstate-to-Hamiltonian'' method, whose central object is the covariance matrix~\cite{qi2017,chertkov2018}. Given a target space of Hamiltonians, the method aims at finding a set of parameters --- a disorder configuration in our case --- minimizing the energy variance $\sigma^2_\mathrm{P}$ of the input state with respect to the new Hamiltonian.

In Sec.~\ref{sec:model_defs}, we define the models and the eigenstate-to-Hamiltonian reconstruction method which simply requires the knowledge of two-point correlators. We then address the simple problem of a one-dimensional Anderson insulator in Sec.~\ref{sec:anderson} for which ground and excited states are investigated. In both cases, a power-law decay of the smallest eigenvalues of the covariance matrix is found, with a disorder-dependent exponent. The associated parent Hamiltonians display sharp spatial features at the special sites where the bipartite entanglement is locally minimal, and spins close to being perfectly aligned or anti-aligned with the local field.
Using numerics and analytical calculations on a toy model, we explore this behavior which, upon increasing the system size results in chain breaks in the thermodynamic limit, thus providing a natural description of the parent Hamiltonians. We further investigate in Sec.~\ref{sec:MBL} the more complex situation of disorder and interaction at high-energy using state-of-the-art exact diagonalization techniques. Quite remarkably, the transition from the ergodic to the so-called many-body localized regime can be captured from the behavior of the smallest non-zero eigenvalue of the covariance matrix: finite in the ergodic phase, and power-law vanishing in the MBL regime. Both phases are analytically understood, with again the remarkable trend for the chain to break in the MBL regime, thus providing a simple picture akin to that of Anderson localization. Finally, we summarize our conclusions in Sec.~\ref{sec:summary_conclusions}. Additional details are provided in Appendexes.

\section{Models and definitions}\label{sec:model_defs}

\subsection{The random-field XXZ models}

In this paper, we focus on the paradigmatic one-dimensional XXZ Hamiltonian describing $L$ interacting spins $S=1/2$ in a local random magnetic field $h_i$, drawn from a uniform distribution $\in[-h,h]$, with $h$ characterizing the disorder strength,
\begin{equation}
    \mathcal{H}_0=\sum_{i=1}^{L-1}S^x_iS^x_{i+1}+S^y_iS^y_{i+1} + \Delta S^z_iS^z_{i+1} +\sum_{i=1}^Lh_i S^z_i.
    \label{eq:H}
\end{equation}
We use open boundary conditions.
This model can be recasted into interacting spinless fermions in a random potential through a Jordan-Wigner transformation, and at $\Delta=0$, the system is equivalent to free fermions which are Anderson localized for any finite disorder strength $h\neq 0$. For later convenience, the above Hamiltonian can be written in the following form, $\mathcal{H}_0=\sum_{i=1}^{L-1}h_{i,i+1} + \sum_{i=1}^L h_iS_i^z$, where the first term represents the pairwise spin-spin couplings acting on the $L-1$ bonds of the open chain, and the second one is a sum of $L$ on-site random field terms. The above model~\eqref{eq:H} has been intensely studied in the past, owing to its localization properties in the ground state~\cite{Giamarchi1988,Bouzerar94,Schmitteckert98,Doggen2017}, as well as more recently as a paradigmatic example for the MBL problem for highly excited states~\cite{nandkishore_many-body_2015,alet_many-body_2018,abanin_many-body_2019}.

\subsection{Eigenstate-to-Hamiltonian construction}

Starting from an input state $\Psi_0$, an eigenstate of the Hamiltonian defined in Eq.~\eqref{eq:H} for a given disorder configuration $\{h_i\}$, we aim at building another parent Hamiltonian $\mathcal{H}_\mathrm{P}$ for $\Psi_0$, by minimizing the energy variance
\begin{equation}
    \sigma^2_\mathrm{P}= \langle\Psi_0|\mathcal{H}^2_\mathrm{P}|\Psi_0\rangle-\langle\Psi_0|\mathcal{H}_\mathrm{P}|\Psi_0\rangle^2,
    \label{eq:var}
\end{equation}
and thus making $\Psi_0$ a good approximation of an actual eigenstate of $\mathcal{H}_\mathrm{P}$. We define and constrain the target space for the possible $\mathcal{H}_\mathrm{P}$ to be of the same form as the original Hamiltonian~\eqref{eq:H}, i.e,
\begin{equation}
    \mathcal{H}_\mathrm{P}=x_0\sum_{i=1}^{L-1}h_{i,i+1}+\sum_{i=1}^L x_i S_i^z,
    \label{eq:HP}
\end{equation}
where the $(L+1)$-dimensional vector $\boldsymbol{x}=(x_0,\ldots,x_L)^T$ contains the new parameters (real numbers) of the parent Hamiltonian~\eqref{eq:HP}. Further simplifying the notations by writing $\mathcal{H}_\mathrm{P}=\sum_{i=0}^{L}x_i{O}_i$ allows us to express the energy variance~\eqref{eq:var} as
\begin{equation}
    \sigma^2_\mathrm{P}=\sum_{i=0}^L\sum_{j=0}^L x_ix_j\Bigl[\langle{O}_i{O}_j\rangle-\langle{O}_i\rangle\langle{O}_j\rangle\Bigr]={\boldsymbol{x}}^T \,\mathsf{C}\,{\boldsymbol{x}},
    \label{eq:cov_mat}
\end{equation}
where $\mathsf{C}$ is the so-called covariance matrix with entries $\mathsf{C}_{ij}=\langle O_iO_j\rangle-\langle O_i\rangle\langle O_j\rangle$, and where the expectation value is taken over the input state $\Psi_0$. From Eq.~\eqref{eq:cov_mat}, it is clear that if $\boldsymbol{x}$ is an eigenvector of $\mathsf{C}$ with zero eigenvalue, the set of parameters contained in $\boldsymbol{x}$ defines a parent Hamiltonian $\mathcal{H}_\mathrm{P}$ for which the initial input state $\Psi_0$ is an exact eigenstate. We note and sort in ascending order the eigenvalues of the covariance matrix, $e_1\leq... e_j...\leq e_{L+1}$, with corresponding eigenvectors $\boldsymbol{x}_j$. Both the total magnetization operator $S^z_\mathrm{tot}=\sum_i S^z_i$ and $\mathcal{H}_0$ itself commute with $\mathcal{H}_0$. Hence, the kernel of $\mathsf{C}$ is two-dimensional with its first two eigenvalues $e_1=e_2=0$. Therefore, we focus on the first non-trivial eigenvalue of the covariance matrix, $e_3$.

\subsection{Covariance matrix}

For the target space of parent Hamiltonians considered in this paper~\eqref{eq:HP}, the covariance matrix entries $\mathsf{C}_{ij}=\langle O_iO_j\rangle_{\Psi_0}-\langle O_i\rangle_{\Psi_0}\langle O_j\rangle_{\Psi_0}$ can be expressed in a simple form, only involving two-body spin correlations. Indeed, using the fact that $O_0=\mathcal{H}_0-\sum_{i=1}^L h_i S_i^z$ and $O_i=S_i^z$ if $i>0$, the matrix elements can be simplified to
\begin{eqnarray}
    \mathsf{C}_{00}&=&\sum_{i=1}^L\sum_{j=1}^Lh_i h_j \Bigl(\langle S_i^z S_j^z\rangle-\langle S_i^z\rangle\langle S_j^z\rangle\Bigr),\nonumber\\
    \mathsf{C}_{0j}&=&\mathsf{C}_{j0}=-\sum_{i=1}^Lh_i\Bigl(\langle S_i^z S_j^z\rangle-\langle S_i^z\rangle\langle S_j^z\rangle\Bigr),\nonumber\\
    \mathsf{C}_{ij} &=&\mathsf{C}_{ji}=\langle S_i^z S_j^z\rangle-\langle S_i^z\rangle\langle S_j^z\rangle~(i\neq j\neq 0),\nonumber\\
    \mathsf{C}_{ii} &=&\frac{1}{4}-\langle S_i^z\rangle^2~(i\neq 0),
    \label{eq:Cij}
\end{eqnarray}
where the expectation value is evaluated over the input state $\Psi_0$. These correlation functions can easily be computed in numerical simulations and are used in practice to evaluate the entries of the covariance matrix $\mathsf{C}$.

\section{Non-interacting Anderson localization}\label{sec:anderson}

We first address the one-dimensional random-field spin-half XY model described by the Hamiltonian~\eqref{eq:H} at $\Delta=0$. Through a Jordan-Wigner transformation, it can be mapped to non-interacting spinless fermions on a chain,
\begin{equation}
    \mathcal{H}_\mathrm{And.}=\sum_{i=1}^{L-1}\frac{1}{2}\Bigl(c^\dag_ic_{i+1}^{\vphantom{\dagger}}+\mathrm{H.c.}\Bigr) +\sum_{i=1}^L h_i c^\dag_ic_i^{\vphantom{\dagger}},
    \label{eq:Anderson}
\end{equation}
where $c^\dag_i$ ($c_i$) is a fermionic creation (annihilation) operator and the random variables $h_i$ act as a disordered potential. Working in the zero magnetization sector in the spin language (half-filling for fermions), we define the energy density above the ground state,
\begin{equation}
    \epsilon=(E-E_\mathrm{min})\Bigl/(E_\mathrm{max}-E_\mathrm{min}),
    \label{eq:epsilon}
\end{equation}
with $E_\mathrm{min}$ and $E_\mathrm{max}$ the extremal eigenenergies of the Hamiltonian describing the system. We focus on $\epsilon=0$ (ground state) as well as $\epsilon=0.5$ (center of the spectrum) in the following.

\subsection{Exact diagonalization results}

\begin{figure}[b!]
    \centering
    \includegraphics[width=.95\columnwidth]{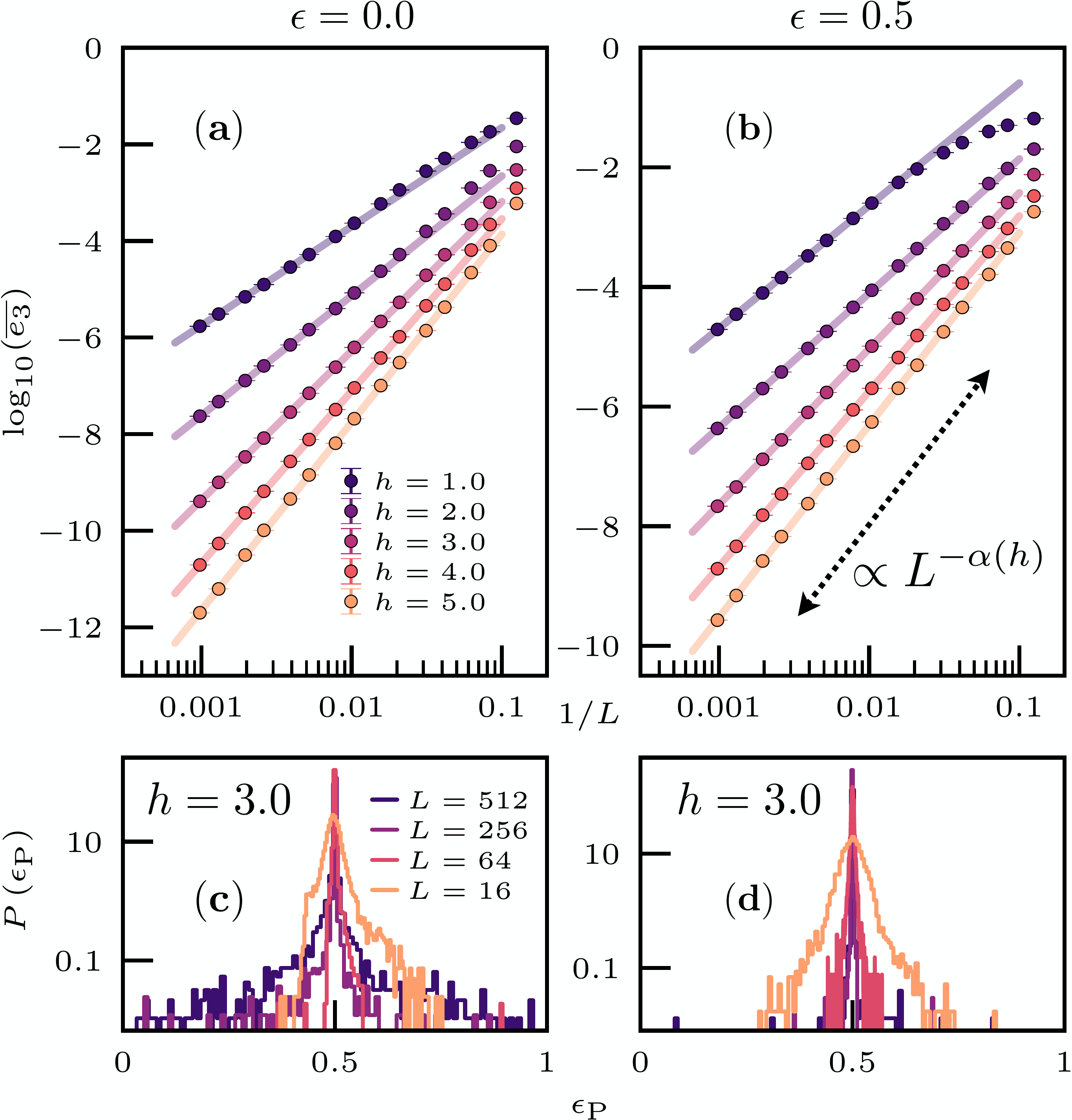}
    \caption{Top: Power-law decay of the disordered average third eigenvalue $\overline{e_3}$ of the covariance matrix, plotted as a function of the inverse system length $1/L$ for $\epsilon=0$ (left) and $\epsilon=0.5$ (right). Various representative disorder strengths $h$ are shown [see also Table~\ref{tab:alpha_anderson} for an estimate of the decay exponents as defined in Eq.~\eqref{eq:e3}]. Bottom: Distribution of the parent energy density $\epsilon_\mathrm{P}$ for input states $|\Psi_\mathrm{loc}\rangle$ at (c) $\epsilon=0$ and (d) $\epsilon=0.5$. Data are shown at $h=3$ for different system sizes $L$.}
    \label{fig:e3anderson}
\end{figure}

\textit{Covariance matrix.---} Once the free-fermion Hamiltonian~\eqref{eq:Anderson} is numerically diagonalized, we readily obtain the spectrum $\{e_i\}_{i=1,\ldots,\,L+1}$ of $\mathsf{C}$ using the substitution $S_i^z=c^\dag_ic_i^{\vphantom{\dagger}}-1/2$ in Eq.~\eqref{eq:Cij}. As already discussed, the first non-trivial non-zero eigenvalue of $\mathsf{C}$ is $e_3$. Its value averaged over $\approx 10^4$ disordered samples is shown in Fig.~\ref{fig:e3anderson}\,(top panels) for a few representative values of the disorder strength $h$. One clearly sees a power-law decay of the form
\begin{equation}
    \overline{e_3}\propto L^{-\alpha(h)},
    \label{eq:e3}
\end{equation}
with a disorder-dependent exponent $\alpha(h)$, growing with $h$ and reported in Table~\ref{tab:alpha_anderson}. All eigenfunctions of~\eqref{eq:Anderson} are localized for any $h\neq 0$. Using such localized eigenstates $|\Psi_\mathrm{loc}\rangle$ as an input states yields a vanishing variance $e_3$ when $L\to +\infty$, a feature already observed for the zero-temperature interacting Bose-glass problem~\cite{dupont2019}. Therefore, we expect $|\Psi_\mathrm{loc}\rangle$ to be a fairly good  approximant of an eigenstate of the associated parent Hamiltonian $\mathcal{H}_\mathrm{P}$, encoded in the corresponding eigenvector $\boldsymbol{x}_3$ of $\mathsf{C}$. This becomes increasingly true for growing disorder strength and system size.

\begin{table}[t!]
    \begin{center}
        \begin{tabular}{l|c|c|c|c|c}
            $h$&1&2&3&4&5\\
            \hline
            $\alpha_{\epsilon=0}$&2.05(1)&2.47(1)&3.10(1)&3.56(1)&3.88(1)\\
            $\alpha_{\epsilon=0.5}$&2.05(1)&2.24(1)&2.61(1)&2.93(1)&3.20(1)\\
            \hline
            \hline
        \end{tabular}
        \caption{    \label{tab:alpha_anderson}
Decay exponent $\alpha(h)$ of the third eigenvalue $e_3$ of the covariance matrix, as defined in Eq.~\eqref{eq:e3}, estimated at energy densities $\epsilon=0$ and $\epsilon=0.5$ for a few representative values of $h$ shown in Fig.~\ref{fig:e3anderson}.}
    \end{center}
\end{table}

\begin{figure*}[ht!]
    \centering
    \includegraphics[width=2\columnwidth]{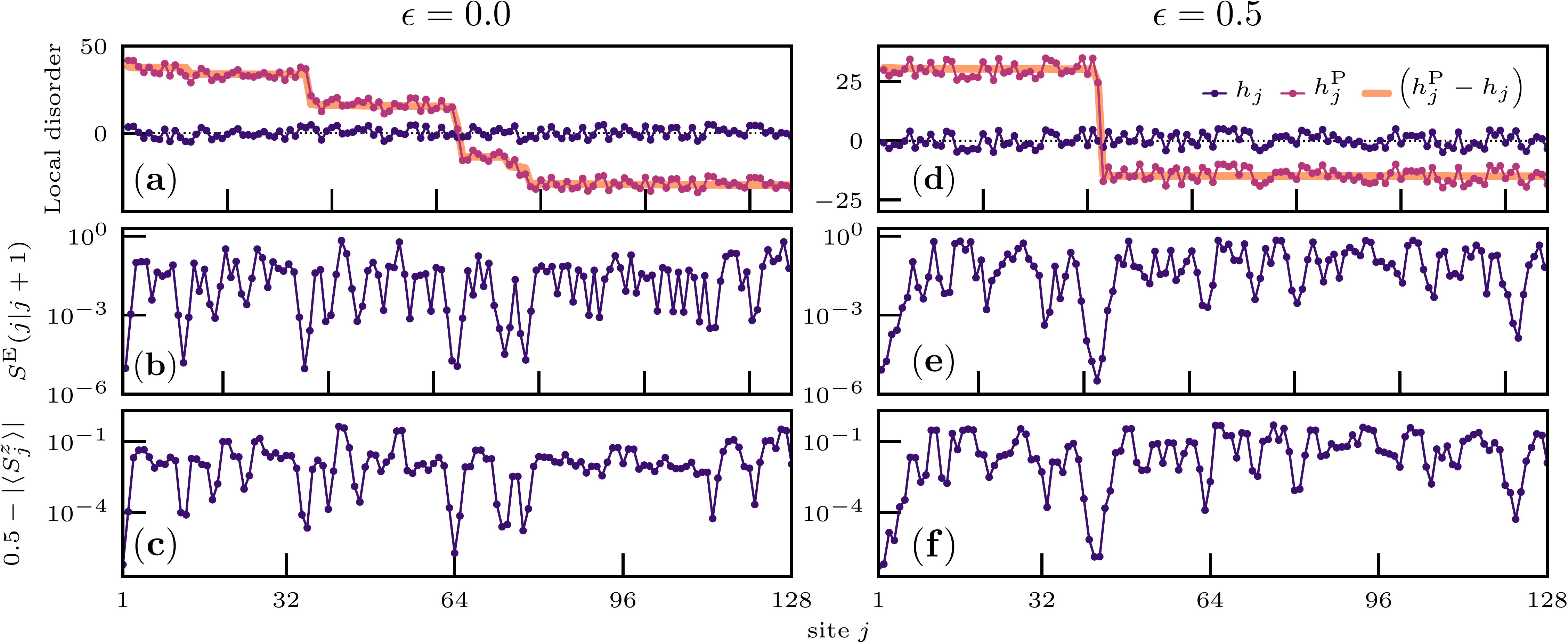}
    \caption{Typical disordered free-fermion chain of $L=128$ sites with $h=5$: ground-state ($\epsilon=0$, left, $e_3=8.33\cdot 10^{-9}$) and highly excited state ($\epsilon=0.5$, right, $e_3=6.05\cdot 10^{-9}$). Top panels (a,\,d): Local distribution in real space of the initial disorder configuration $\{h_i\}$ (purple) compared to the new configuration $\{h^\mathrm{P}_i\}$ (red), together with their difference (orange lines). Middle panels (b,\,e): Entanglement entropy $S^\mathrm{E}(i|\,i+1)$ between subsystems $[1,\,i]$ and $[i+1,\,L]$. Bottom panels (c,\,f): Local magnetization shown as a deviation to the full polarization.}
    \label{fig:hlocal_anderson}
\end{figure*}

\textit{Parent Hamiltonians.---} The first relevant feature concerns the parent energy associated with $|\Psi_\mathrm{loc}\rangle$,
\begin{equation}
    E_\mathrm{P}=\langle\Psi_\mathrm{loc}|\mathcal{H}_\mathrm{P}|\Psi_\mathrm{loc}\rangle.
\end{equation}
In terms of energy density measured above the ground-state, as defined above in Eq.~\eqref{eq:epsilon}, we find that the parent energy is  sharply distributed around $\epsilon_\mathrm{P}\sim 0.5$, as shown in Fig.~\ref{fig:e3anderson}\,(bottom panels), regardless of the initial value of $\epsilon$ (zero or one half). This reflects the peak at $\epsilon\sim 0.5$ in the underlying density of states. Note, however, that finite-size effects are distinct between the [Fig.~1\;(c)] ground and [Fig.~1\;(d)] excited input states.

By construction, the set of parent Hamiltonians has the same form as the original one [see Eq.~\eqref{eq:HP}], but with a new random field configuration $\{h^\mathrm{P}_i\}$. It is enlightening to compare it with the initial one $\{h_i\}$, as done in Fig.~\ref{fig:hlocal_anderson}\,(top panels) for a typical sample of size $L=128$ at $h=5$. Interestingly, they are strongly correlated, such that over finite segments $s$ we observe $h^\mathrm{P}_i \simeq h_i+\mathcal{C}_s$. In other words, the parent field is simply shifted by a global constant $\mathcal{C}_s$.
It thus forms a plateau-type structure, with sharp steps between consecutive segments, of amplitude much larger than the average disorder strength $h$.

Interestingly, the jump positions $i$ correspond precisely to minima in the entanglement entropy between subsystems $A=[1,\,i]$ and $B=[i+1,\,L]$,
\begin{equation}
    S^\mathrm{E}(i|\,i + 1)=-\mathrm{Tr}_B\Bigl[{\hat{\rho}}_B\,\log {\hat{\rho}}_B\Bigr],
\end{equation}
where $\hat{\rho}_B=\mathrm{Tr}_A|\Psi_\mathrm{loc}\rangle\langle\Psi_\mathrm{loc}|$. This is visible in the middle panels of Fig.~\ref{fig:hlocal_anderson}.
At these positions, the chain is cut into two almost independent pieces.
Furthermore, entanglement minima are located at those sites where spins are very close to being perfectly polarized, as visible in the lower panels of  Fig.~\ref{fig:hlocal_anderson}.
We further discuss the physics of spin polarization in the next section.

\subsection{Disorder-induced spin polarizations}

\begin{figure}[t!]
    \centering
    \includegraphics[width=\columnwidth]{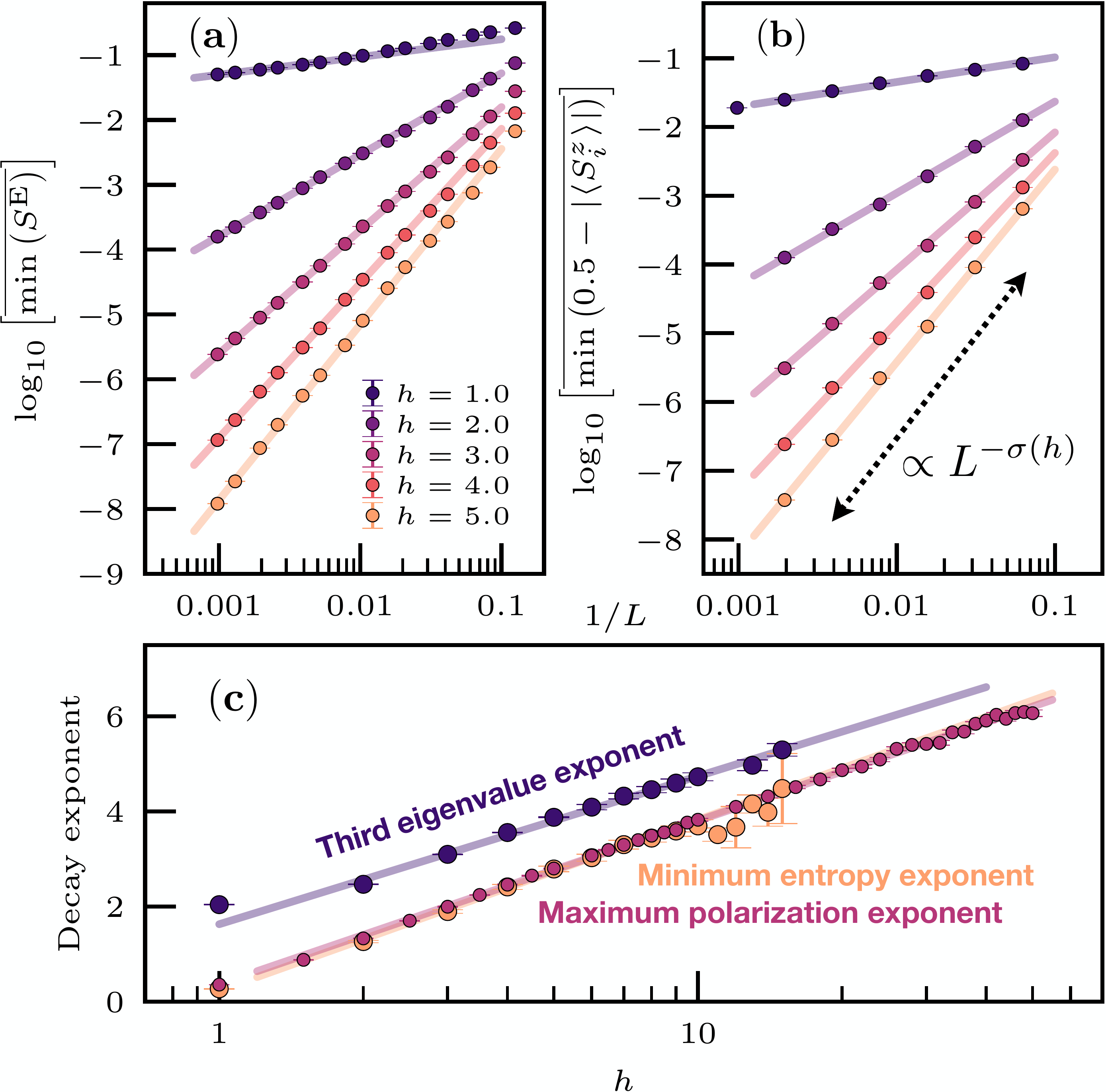}
    \caption{Exact diagonalization results for both (a) the minimal entanglement entropy and (b) the maximal polarization, plotted as a function of the inverse chain length $1/L$. Power-law decay is clearly visible for a few representative values of the disorder strength $h$, as indicated on the plot. Panel (c) shows the disorder dependence of the exponents, in a log-linear scale. Straight lines are fits of the form $a+b\log h$ with $a=0.23(4)$, $0.37(3)$, $1.63(7)$ and $b=1.56(3)$, $1.49(1)$, $1.35(4)$, respectively, for entanglement, polarization, and $e_3$.}
    \label{fig:exponents_anderson}
\end{figure}

\textit{Numerics.---} Without loss of generality, the discussion is done for the case $\epsilon=0$ here. Comparable results are observed for $\epsilon>0$, as discussed in Appendix~\ref{app:distributions_Anderson}. We first focus on the maximally (or minimally) occupied site, such that in spin language, the quantity $1/2-|\langle S_i^z\rangle|$ (bottom panels of Fig.~\ref{fig:hlocal_anderson}) is minimized along the chain. Interestingly, this quantity vanishes algebraically with increasing system size, as shown in Fig.~\ref{fig:exponents_anderson}\,(b),
\begin{equation}
    \min_i \left[ 1/2-{\overline{|\langle S_i^z\rangle|}} \right] \propto L^{-\gamma(h)}.
\end{equation}
This result implies that disorder will cut the chain in the thermodynamic limit. Consequences for the entanglement across such ``bottlenecks'' are also very interesting, as displayed in Fig.~\ref{fig:exponents_anderson}\,(a) where the minima $\mathrm{min}(S^\mathrm{E})$ also decays with a similar power-law. Indeed, considering a two-site system (A-B) with a strongly polarized moment, $1/2-|\langle S^z_\mathrm{A}\rangle |\sim L^{-\gamma}$, for any $\langle S^z_\mathrm{B}\rangle$ a straightforward calculation yields
\begin{equation}
    S^\mathrm{E}\propto L^{-\gamma}\langle S^{z}_\mathrm{B}\rangle^2.
    \label{eq:SEgamma}
\end{equation}
Such similar power-law scalings for both minimal entropy and maximal polarization are clearly visible in panels (a-b) of Fig.~\ref{fig:exponents_anderson} as a function of $1/L$, together with the exponents in panel (c) where a very good agreement between entropy and polarization exponents is observed. Additionally, the third eigenvalue exponent $\alpha(h)$, from
Eq.~\eqref{eq:e3} displays a similar disorder dependence, which turns out to be non-trivial $\sim \log h$, see the caption of Fig.~\ref{fig:exponents_anderson}.

\begin{figure}[t!]
    \centering
    \includegraphics[width=.9\columnwidth]{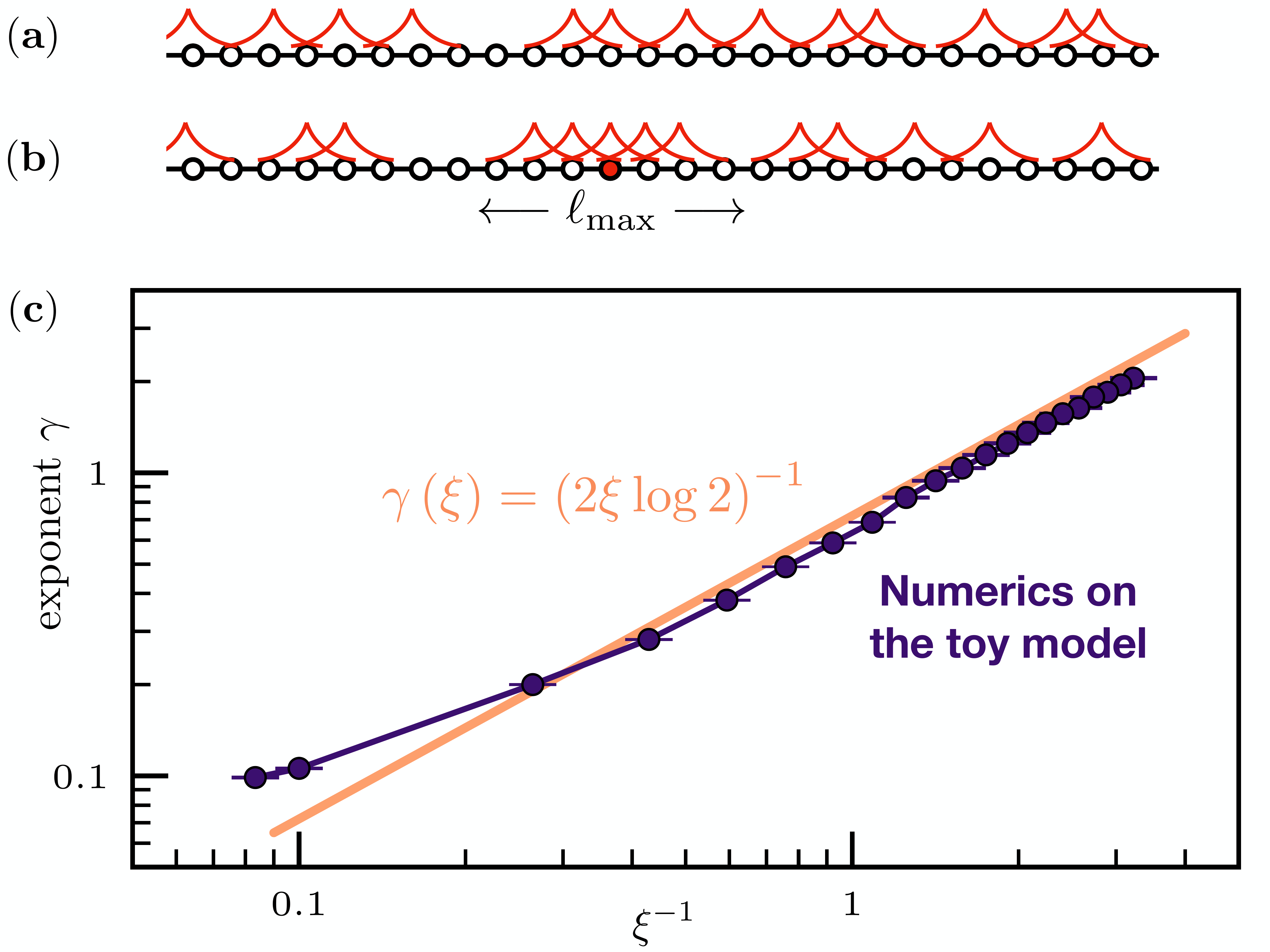}
    \caption{(a-b) Schematic representation of a collection of one-dimensional localized orbitals of the form~\eqref{eq:loc}. (a) Generic case for a random state. In (b) we show the maximally occupied site (red circle), a situation realized when a region of  $\ell_\mathrm{max}$ consecutive sites are occupied. (c) Decay exponent governing the deviation from the full polarization of the maximally occupied site computed from a numerical simulation of the toy model at half-filling (purple symbols) compared to the analytical prediction of Eq.~\eqref{eq:gamma_xi} (orange line).}
    \label{fig:orbitals}
\end{figure}

\textit{Toy model.---} To get an analytical understanding of this behavior, we build a toy model which provides a simplified picture for free fermions. Consider a collection of one-dimensional localized orbitals, as schematized on Fig.~\ref{fig:orbitals}, of the form
\begin{equation}
    |\phi_k(i)|^2=A_k\exp\left(-\frac{|i-i_0^k|}{\xi_k}\right)\quad (i=1,\ldots,L),
    \label{eq:loc}
\end{equation}
where $k$ labels the orbital, $\xi_k$ is the associated localization length,  $i_0^k$ is the localization center, and $A_k=\tanh(1/2\xi_k)$ a normalization factor. The particle density at site $i$ is given by the sum of occupied orbitals
\begin{equation}
    n(i)=\sum_{k\,\mathrm{occcupied}}|\phi_k(i)|^2.
\end{equation}
We will assume for simplicity that all orbitals have the same localization length $\xi_k\equiv\xi$. Within such a toy model, the maximally occupied site with $n_\mathrm{max}=\max_{i}\left[n(i)\right]$ is expected when $\ell_\mathrm{max}$ consecutive sites are occupied, as schematized in Fig.~\ref{fig:orbitals}\,(b). At half filling, a configuration with $\ell$ consecutive sites occupied occurs with probability $P(\ell)\approx 2^{-\ell}$. In the large system size limit, the longest region has probability $\propto 1/L$ such that
\begin{equation}
    \ell_\mathrm{max}\approx \log L/\log 2.
\end{equation}
Therefore, the maximal occupation is very close to one,
\begin{eqnarray}
    n_\mathrm{max}&=&\tanh\left(\frac{1}{2\xi}\right)\Bigl(1+2\sum_{r=1}^{\ell_\mathrm{max}/2}\mathrm{e}^{-r/\xi}\Bigr)\nonumber\\
    &=&1-\frac{\exp\left(-\frac{1}{2\xi}\right)}{\cosh\left(\frac{1}{2\xi}\right)}L^{-\gamma(\xi)},
    \label{eq:nmax}
\end{eqnarray}
with a decay exponent
\begin{equation}
    \gamma(\xi)\approx\left(2\xi\log 2\right)^{-1}.
    \label{eq:gamma_xi}
\end{equation}
This analytical expression can be checked against numerical simulations of the toy model, as shown in Fig.~\ref{fig:orbitals}\,(c) where one sees a very good agreement for the exponent $\gamma(\xi)$ with Eq.~\eqref{eq:gamma_xi}.

Coming back to the real Anderson model, at strong random field strength $h\gg 1$, the disorder dependence of the localization length is easy to obtain. Indeed, a perturbative expansion of any wave function away from its localization center gives an amplitude vanishing $\sim h^{-2r}$, where $r$ is the distance to the localization center. This yields $\xi=(2\log h)^{-1}$, and thus
\begin{equation}
    \gamma(h)\approx\log h/\log 2,
\end{equation}
in good agreement with exact diagonalization results displayed in Fig.~\ref{fig:exponents_anderson}\,(c) where the decay exponent is well described by the form $a+b\log(h)$, with $b\sim 1.5$, which is quite close to the predicted $1/\log 2\approx 1.44$.

\subsection{Consequences for the covariance matrix and the parents Hamiltonians}
\label{subsec:consequences}
If a site $i_0$ were fully polarized, some entries in $\mathsf{C}$, given by Eq.~\eqref{eq:Cij}, would vanish: $\mathsf{C}_{i_0,j}=0$ $\forall j$, implying that $e_3$ would vanish as well, and that the parent Hamiltonian would decompose into two disconnected parts $i<i_0$ and $i>i_0$.
However, such a chain breaking only occurs in the thermodynamic limit, as seen before.
For finite systems, the diagonal entries are power-law vanishing $\mathsf{C}_{i_0,i_0}\propto L^{-\gamma}$ at maximally polarized sites.

\begin{figure}[b!]
    \centering
    \includegraphics[width=\columnwidth,clip]{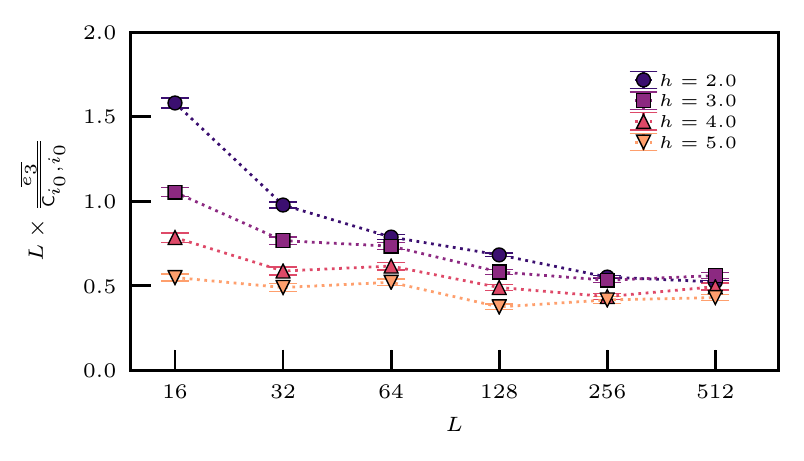}
    \caption{Ratio, as defined Eq.~\eqref{eq:ratio}, between the average smallest non-zero eigenvalue $\overline{e_3}$ and the average smallest diagonal element of the covariance matrix $\overline{\mathsf{C}_{i_0,i_0}}$ rescaled by the system length $L$, plotted against $L$ for different disorder strengths $h$. Exact diagonalization results for the one-dimensional Anderson model~\eqref{eq:Anderson} at $\epsilon=0$. We clearly see convergence toward a finite value.}
    \label{fig:ratio}
\end{figure}

We now argue that the power-law decay of the diagonal entries dictates the power-law decay of $e_3$.
For free-fermions, all connected correlators of the form $\langle S_i^z S_j^z\rangle-\langle S_i^z\rangle\langle S_j^z\rangle$ are negative if $i\neq j$ and positive if $i=j$.
Therefore, one can interpret the covariance matrix as a tight-binding Hamiltonian whose negative off-diagonal elements are kinetic terms favoring delocalization of the wave functions: One can write $e_3(L)=\mathsf{C}_{i_0,i_0}(L)-\left|t^\mathrm{eff}_{i_0}(L)\right|$ where $\left|t^\mathrm{eff}_{i_0}(L)\right|$ accounts for the effective ``kinetic energy'' gain.
It remains an open problem to understand the precise size dependence of $\left|t^\mathrm{eff}_{i_0}(L)\right|$.
We resort to numerics, and observe in Fig.~\ref{fig:ratio} that
\begin{equation}
    L\times\overline{e_3}\Bigl/\overline{\mathsf{C}_{i_0,i_0}} \sim \text{cst},
    \label{eq:ratio}
\end{equation}
yielding $\left|t^\mathrm{eff}_{i_0}(L)\right|\propto \mathsf{C}_{i_0,i_0}\left(1 - 1/L\right)$ and $e_3(L)\propto L^{-\gamma-1}$.
These results highlight the link between the spin polarization induced by locally strong disorder and the power-law decay of the minimum non-zero eigenvalue $e_3$. In the thermodynamic limit, perfect polarization occurs for a set of sites $\{i_0\}$. By the previous reasoning, this implies the vanishing of a corresponding set of $e_{j\geq 3}$, and the existence of an associated family of exact parent Hamiltonians. In Appendix~\ref{app:alpha_i}, we numerically confirm this fact in the case of an MBL system.

\section{Many-body localization}\label{sec:MBL}

\begin{figure}[!t]
    \centering
    \includegraphics[width=0.8\columnwidth]{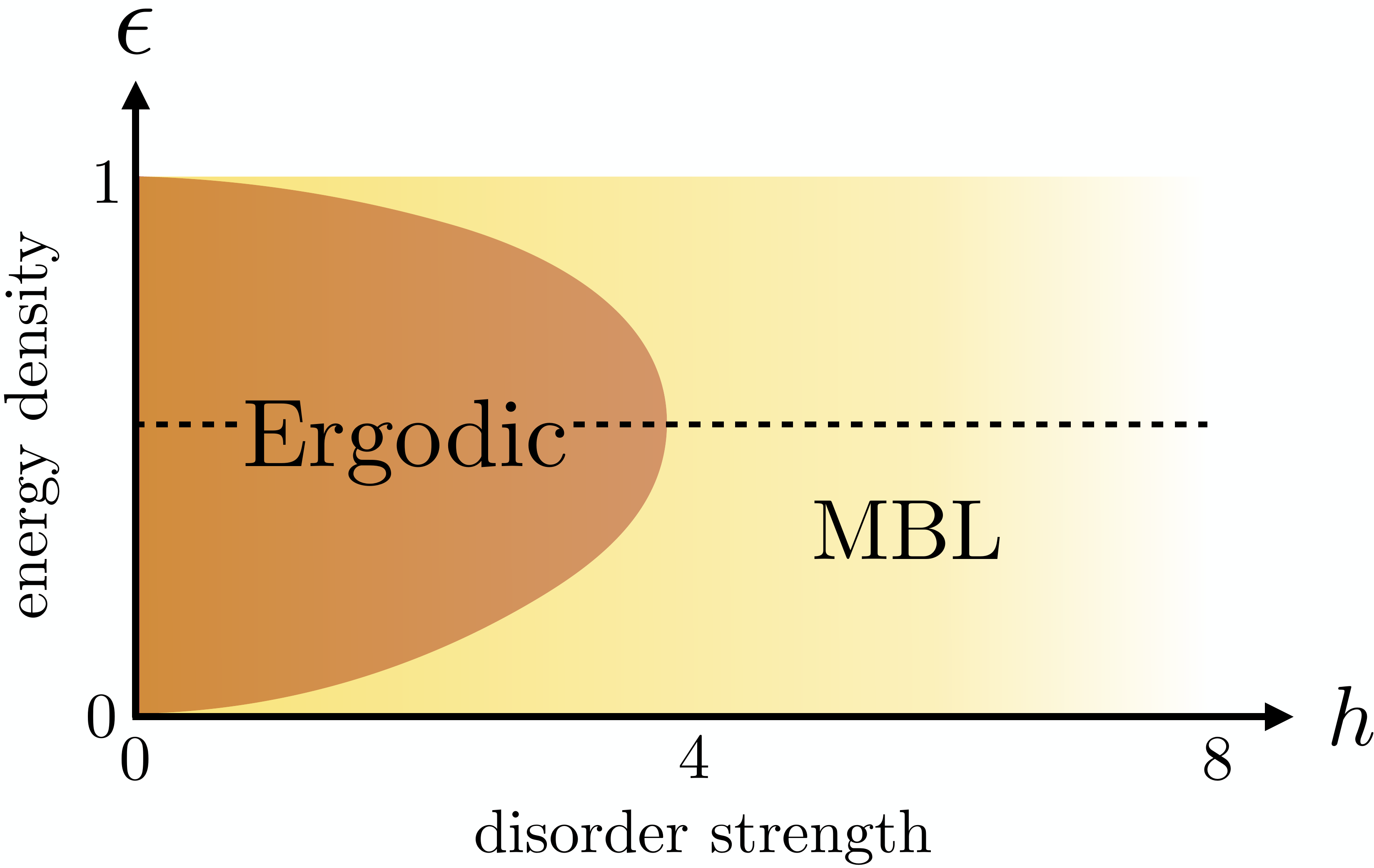}
    \caption{Schematic picture for the disorder $h$ --- energy density $\epsilon$ phase diagram of the random-field Heisenberg chain model defined in Eq.~\eqref{eq:H} at $\Delta=1$. In this work, the transition is studied for eigenstates in the middle of the energy spectrum, $\epsilon\sim 0.5$, along the dashed line.}
    \label{fig:mbl}
\end{figure}

We now turn our attention to the random-field Heisenberg chain, Eq.~\eqref{eq:H} with $\Delta=1$. This model is well known to exhibit a high-energy eigenstate transition as a function of the disorder strength~\cite{pal_many-body_2010,luitz_many-body_2015,alet_many-body_2018} between a thermal ergodic phase for $h<h_c$ and a non-thermal MBL regime for $h\ge h_c$. The disorder strength versus energy density phase diagram obtained in Ref.~\onlinecite{luitz_many-body_2015} is schematized in Fig.~\ref{fig:mbl}. In the following we focus on the middle of the many-body spectrum, $\epsilon=0.5$, where exact numerical methods give a critical disorder strength $h_c\sim 3.8$~\cite{luitz_many-body_2015,mace_multifractal_2018}. So far, this dynamical transition has been captured using various observables, e.g., level statistics~\cite{jacquod_emergence_1997,pal_many-body_2010,luitz_many-body_2015,serbyn_spectral_2016}, entanglement entropy~\cite{bauer_area_2013,kjall_many-body_2014,luitz_many-body_2015,yu_bimodal_2016}, inverse participation ratio~\cite{luca_ergodicity_2013,luitz_many-body_2015,mace_multifractal_2018,Pietracaprina_2019}, or out-of-equilibrium dynamics~\cite{znidaric_many-body_2008,bardarson_unbounded_2012,serbyn_universal_2013,andraschko_purification_2014,naldesi_detecting_2016,luitz_extended_2016,khait_transport_2016,znidaric_diffusive_2016,doggen_many-body_2018}.
Here, building on the eigenstate-to-Hamiltonain construction, we propose to shed light on this exotic transition.

\subsection{Eigenstate-to-Hamiltonian construction across the ergodic-MBL transition}

In the case of non-interacting localized fermions, we concluded for the existence of parent Hamiltonians for any given \textit{localized} eigenstate. Similar conclusions were also reached for localized ground-states in the presence of an interaction, namely, in the Bose-glass state~\cite{dupont2019}. However, the situation has been shown~\cite{qi2017,garrison2018} to be different for delocalized systems satisfying the eigenstate thermalization hypothesis (ETH)~\cite{deutsch_quantum_1991,srednicki_chaos_1994,dalessio_quantum_2016}. In such a case, the Hamiltonian is expected to be uniquely defined from a given eigenstate. In the language of eigenstate-to-Hamlitonian construction, this means that the third smallest eigenvalue $e_3$ of the covariance matrix should remain finite even in the thermodynamic limit.\\

\begin{figure}[t!]
    \centering
    \includegraphics[width=\columnwidth]{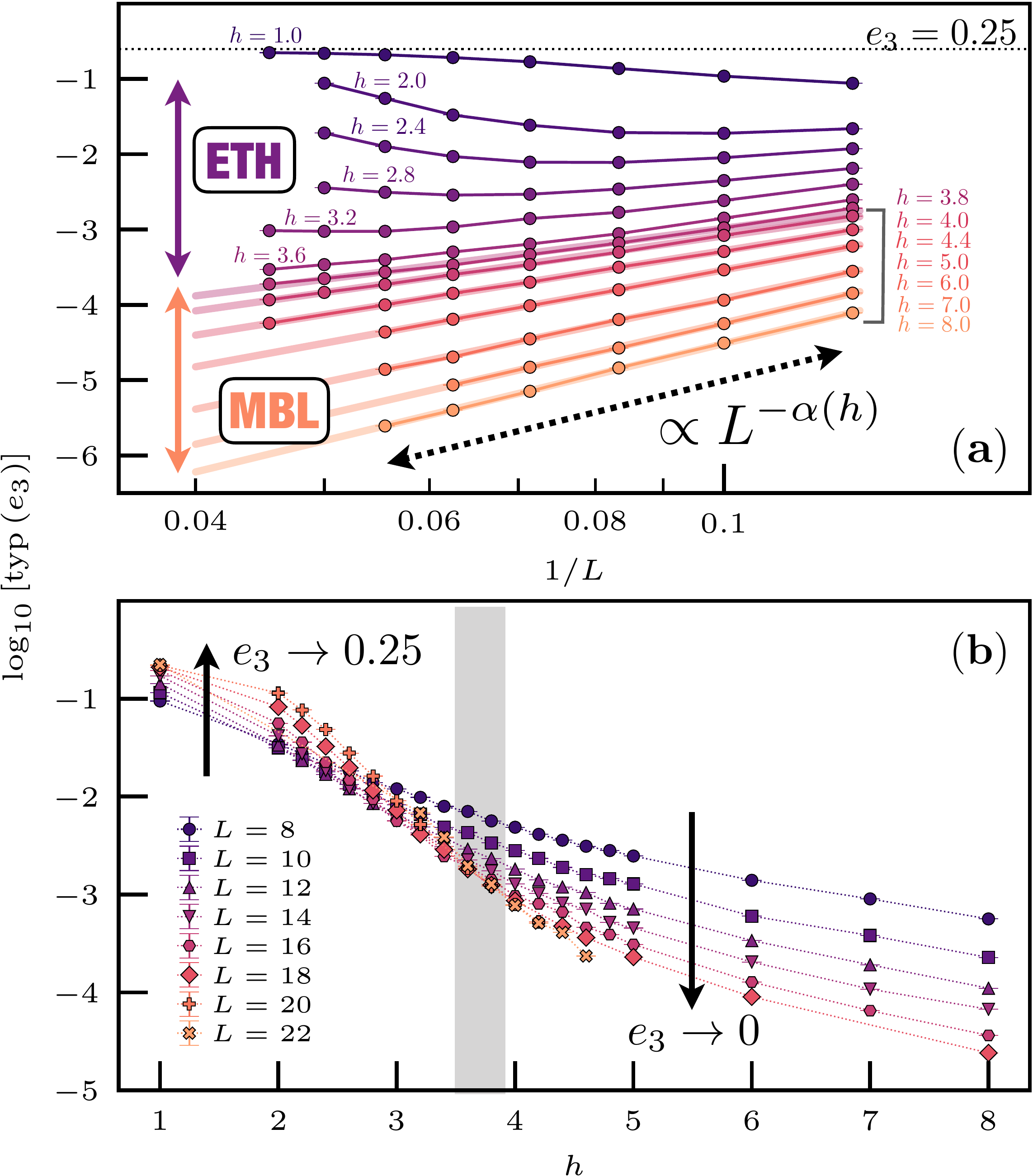}
    \caption{First non-zero eigenvalue $e_3$ of the covariance matrix $\mathsf{C}$ for the random-field Heisenberg chain of Eq.~\eqref{eq:H} at $\Delta=1$. Both panels show the typical value $\overline{\log_{10} e_3}$. (a) Finite size scaling vs $1/L$: Log-log plot showing the decay $L^{-\alpha(h)}$ in the MBL regime for $h\ge 3.8$, and the convergence to a finite value in the ETH phase. (b) Same data plotted against disorder strength $h$ show a crossing separating the two different regimes with a pronounced drift at small sizes. The data corresponds to shift-invert exact diagonalization results obtained in the middle of the spectrum at $\epsilon=0.5$, with the average performed over at least $5\,000$ random samples, using one to ten eigenstate(s) per sample.}
    \label{fig:scaling_e3_mbleth}
\end{figure}

\textit{Scaling of $e_3$.---} We numerically explore the ETH-MBL transition of the random-field Heisenberg chain of Eq.~\eqref{eq:H} at $\Delta=1$. In Fig.~\ref{fig:scaling_e3_mbleth} the behavior of the third eigenvalue $e_3$ of $\mathsf{C}$ is shown as a function of disorder and system size (here again $e_1=e_2=0$ due to total energy and total magnetization conservation). For such a high-energy interacting problem, we can no longer rely on free-fermion methods, and turn to exact diagonalization shift-invert techniques~\cite{luitz_many-body_2015,pietracaprina2018} to deal with the exponentially growing Hilbert space, allowing us to reach systems up to $L=22$ spins. In Fig.~\ref{fig:scaling_e3_mbleth}\,(a) the finite size scaling $e_3(L)$ clearly shows two qualitatively different behaviors: In the MBL regime a power-law decay $e_3 \propto L^{-\alpha(h)}$ akin to that of the Anderson localization is observed, while in the ergodic regime $h<h_c$, $e_3$ remains finite. This striking difference is highlighted in Fig.~\ref{fig:scaling_e3_mbleth}\,(b) where  one observes a crossing in the vicinity of $h_c$. However, we note strong finite size effects with a pronounced drift for small sizes.

\begin{figure}
    \centering
    \includegraphics[width=0.82\columnwidth]{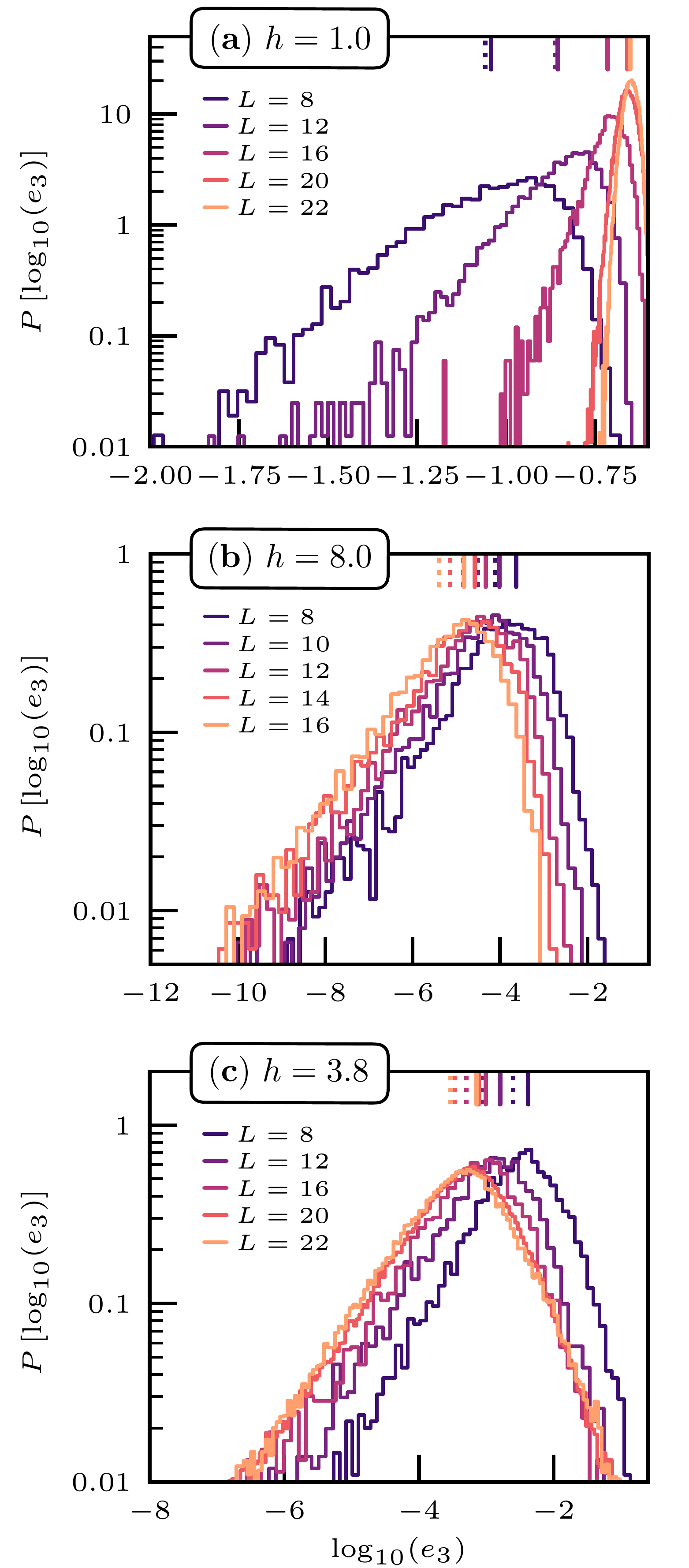}
    \caption{Histograms $P\left[\log_{10}(e_3)\right]$ collected over at least $5\,000$ random samples, using one to ten eigenstate(s) per sample, for three representative disorder strengths: (a) $h=1$ in the ETH regime, (b) $h=8$ MBL, and (c) $h=3.8$ close to the transition. Self-averaging is observed in (a), with a fast shrinking of the distributions with increasing $L$, and typical (solid lines) and average (dotted lines) which gradually coincide. Instead, broad and non-self-averaging distributions are observed in panels (b-c) with a clear separation between typical and average values.}
    \label{fig:distrib_e3_mbleth}
\end{figure}

\textit{Distribution of $e_3$.---} Before discussing the asymptotic forms in both regimes, let us address how $e_3$ is distributed across random samples. This is shown in Fig.~\ref{fig:distrib_e3_mbleth} for three typical disorder strengths. While in the (a) ETH regime one observes a fast shrinking with system size toward the average $e_3\to 1/4$, the situation is radically different for both (b) MBL and (c) at the transition, where one observes very broadly distributed $e_3$, the absence of self-averaging, and a clear distinction between typical and average values (see also Appendix~\ref{app:distributions_MBL} for strong disorder distributions). In order to avoid abnormal rare events, we focus on the typical value $\overline{\log_{10}(e_3)}$ in Fig.~\ref{fig:scaling_e3_mbleth}, instead of the average one. Note that we did not need to consider the typical value in the Anderson localization case since the typical and average values coincide with one another (see Appendix~\ref{app:distributions_Anderson}). Indeed, rare events are less pronounced for Anderson than for MBL, for which there will always exist, albeit small, thermal subregions in the system.

\begin{figure}[h!]
    \centering
    \includegraphics[width=\columnwidth]{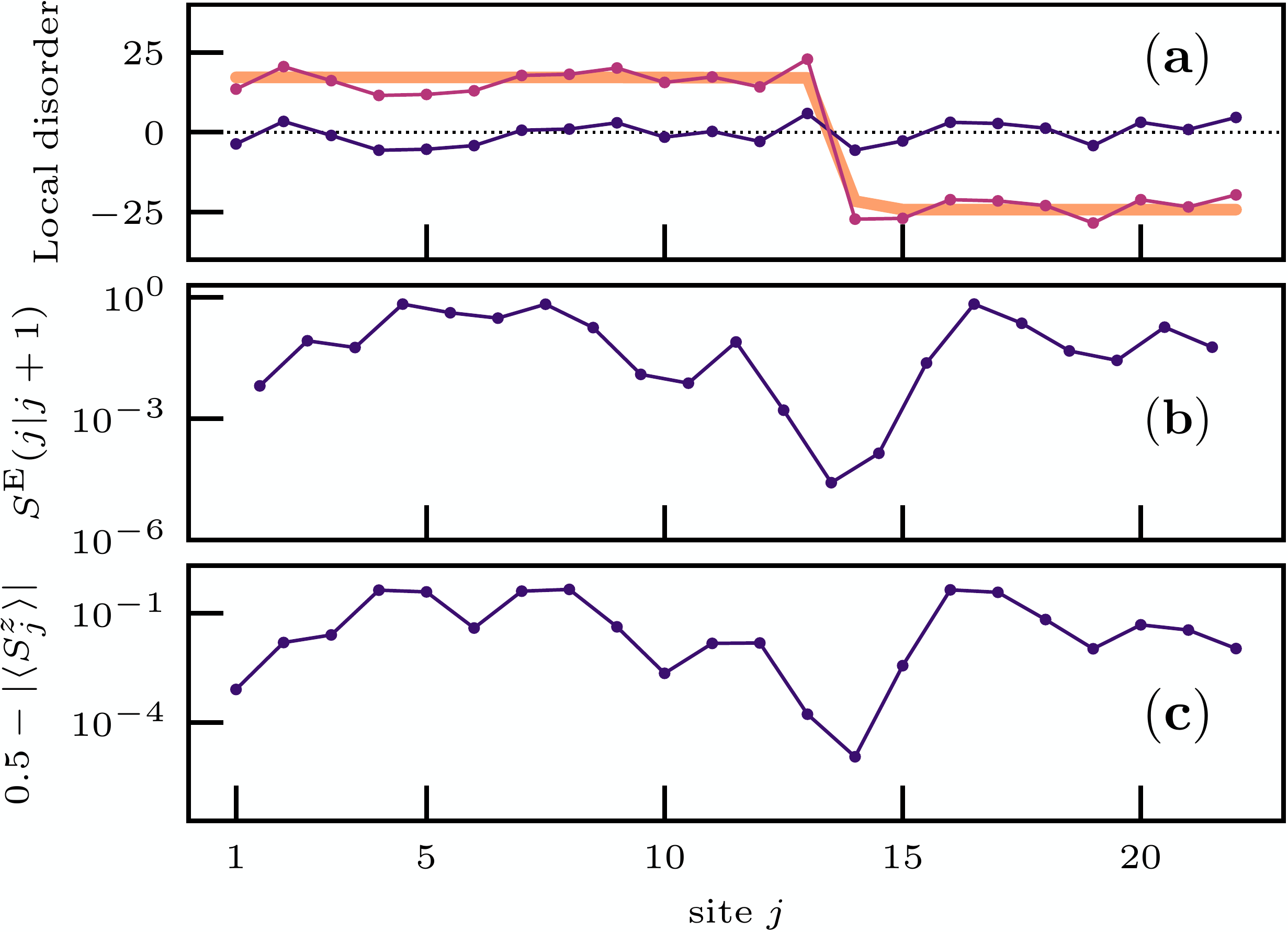}
    \caption{Same as Fig.~\ref{fig:hlocal_anderson} shown here for the interacting case in the MBL regime at $h=6$ for a high-energy eigenstate of an $L=22$ sites chain ($\epsilon=0.5$, $e_3=3.26\cdot 10^{-7}$). (a) Local distribution in real space of the initial disorder configuration $\{h_i\}$ (purple) compared to the new configuration $\{h^\mathrm{P}_i\}$ (red), together with their difference (orange lines). (b) Entanglement entropy $S^\mathrm{E}(i|\,i + 1)$ between subsystems $[1,\,i]$ and $[i+1,\,L]$. (c) Local magnetization shown as a deviation to the full polarization. The jump in the parent field $\{h^\mathrm{P}_i\}$ (a) is precisely observed at the minimum of entanglement (b), corresponding to the maximally polarized site (c).}
    \label{fig:parentH_MBL}
\end{figure}

\textit{Parent MBL Hamiltonians.---} Similarly to Anderson localization, MBL eigenstates are very good approximations for eigenstates of parent Hamiltonians having the same form, only differing with their local random-field configuration. Figure~\ref{fig:parentH_MBL} shows an example for a given MBL eigenstate of $\mathcal{H}_0$, similarly to Fig.~\ref{fig:hlocal_anderson}. Again, a step-like structure is observed, with a perfect correlation of the fields, and the jump occurring at the entropy minimum, also corresponding the maximally polarized spin.

\begin{figure}[t!]
    \centering
    \includegraphics[width=\columnwidth]{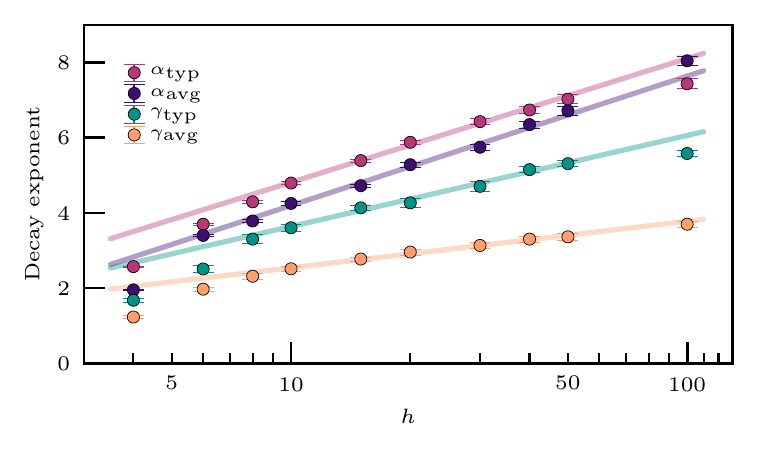}
    \caption{Strong disorder dependence in a log-linear scale of the exponents $\alpha$ and $\gamma$ governing the decays of $e_3(L)$~\eqref{eq:e3avgtyp} and the maximum polarization~\eqref{eq:mzavgtyp}. Both average and typical values are shown. Straight lines are fits of the form $a+b\log h$ with $b_\mathrm{avg}=1.49(8)$, $b_\mathrm{typ}=1.43(4)$ and $a_\mathrm{avg}=0.8(3)$, $a_\mathrm{typ}=1.5(2)$ for $\alpha(h)$. The parameters for  $\gamma(h)$ are $b_\mathrm{avg}=0.54(3)$, $b_\mathrm{typ}=1.05(5)$ and $a_\mathrm{avg}=1.3(8)$, $a_\mathrm{typ}=1.2(2)$.}
    \label{fig:exponents_MBL}
\end{figure}

\textit{Maximally polarized sites.---} In the spirit of our previous findings for Anderson localized states, we also identify for MBL states a chain-breaking mechanism through which parent Hamiltonians become exact in the thermodynamic limit. Indeed, in the MBL regime there are sites where the local magnetization $\langle S_i^z\rangle$ is very close to perfect polarization. At these sites, a step occurs in the parent Hamiltonians' local term, and the associated $e_3$ eigenvalue is very small. While finite size scaling cannot be performed over orders of magnitude such as for free fermions, we still can extract both exponents $\alpha(h)$ and $\gamma(h)$ between $L=8$ and $L=22$, see Fig.~\ref{fig:scaling_e3_mbleth}\,(a). We further note that our analysis is performed very deep in the MBL regime where finite-size effects are not very strong, such that the extracted exponents are reliable. These exponents govern the decay of the energy variance,
\begin{eqnarray}
    \overline{e_3}&\propto& L^{-\alpha_\mathrm{avg}(h)},\nonumber\\
    \exp\left(\overline{\log e_3}\right)&\propto&  L^{-\alpha_\mathrm{typ}(h)},
    \label{eq:e3avgtyp}
\end{eqnarray}
and the maximal polarization
\begin{eqnarray}
    \frac{1}{2}-\left|\langle S_i^z\rangle\right|_\mathrm{max}&\propto&  L^{-\gamma_\mathrm{avg}(h)},\nonumber\\
    \exp\left[{\overline{\log \left(\frac{1}{2}-\left|\langle S_i^z\rangle\right|_\mathrm{max}\right)}}\right]&\propto&  L^{-\gamma_\mathrm{typ}(h)}.
    \label{eq:mzavgtyp}
\end{eqnarray}
The above expressions define average and typical exponents, which are shown in Fig.~\ref{fig:exponents_MBL} as a function of disorder strength. Interestingly, we also find here a logarithmic scaling of the exponent with disorder strength, with a distinction between average and typical values, which is a direct consequence of the absence of self-averaging for local observables, as discussed in Appendix~\ref{app:distributions_MBL}. This distinction is more pronounced for $\gamma(h)$ than for $\alpha(h)$. Indeed, the numerics is compatible with an identical scaling $\alpha_\text{typ}(h)\approx \alpha_\text{avg}(h)\sim 1.45\log h$ of the variance exponent at strong disorder. For the polarization, however, we find $\gamma_\text{typ}(h)/\gamma_\text{avg}(h)\simeq 2$ at strong disorder. Those different behaviors can be related to the differences in the distributions of these two quantities, as discussed in Appendix~\ref{app:distributions_MBL}. In contrast, Anderson localization does not show such a distinction between average and typical exponents (see Appendix~\ref{app:distributions_Anderson}).

\subsection{Analytical results and microscopic picture}

\textit{Ergodic regime.---} Before discussing the MBL regime and the disorder dependence of the decay exponents, let us first address the peculiar behavior observed in the ETH regime. In the ETH phase at infinite temperature, assuming perfect thermalization, the covariance matrix elements given by Eq.~\eqref{eq:Cij} are traces of local operators. They read,
\begin{eqnarray}
	\mathsf{C}_{00} &=& \frac{1}{4} \left[ \sum_{i=1}^N h_i^2 - \frac{1}{N-1} \sum_{i \neq j} h_i h_j \right],\nonumber\\
	\mathsf{C}_{0j} &=& \frac{1}{4} \left[ h_j - \frac{1}{N-1} \sum_{i \neq j} h_i \right],\nonumber\\
	\mathsf{C}_{i\neq 0,j\neq 0} &=& \frac{1}{4} \left[ \delta_{ij} - \frac{1-\delta_{ij}}{N-1} \right].
    \label{eq:ETH}
\end{eqnarray}
One finds the two zero eigenvalues, expected from symmetry considerations. The following $N-2$ values are degenerate and equal to $N/4(N-1)$. Finally, the last one is extensive. This calculation gives an upper bound on the covariance eigenvalues, as it assumes perfect thermalization, which never occurs in a finite system. In the ETH phase, we thus expect $e_3 \leq N/4(N-1)$. In the thermodynamic limit, the system thermalizes completely and we predict $e_3 \to 1/4$, as we nicely observe in Fig.~\ref{fig:scaling_e3_mbleth}.

\begin{figure}[t!]
    \centering
    \includegraphics[width=0.7\columnwidth]{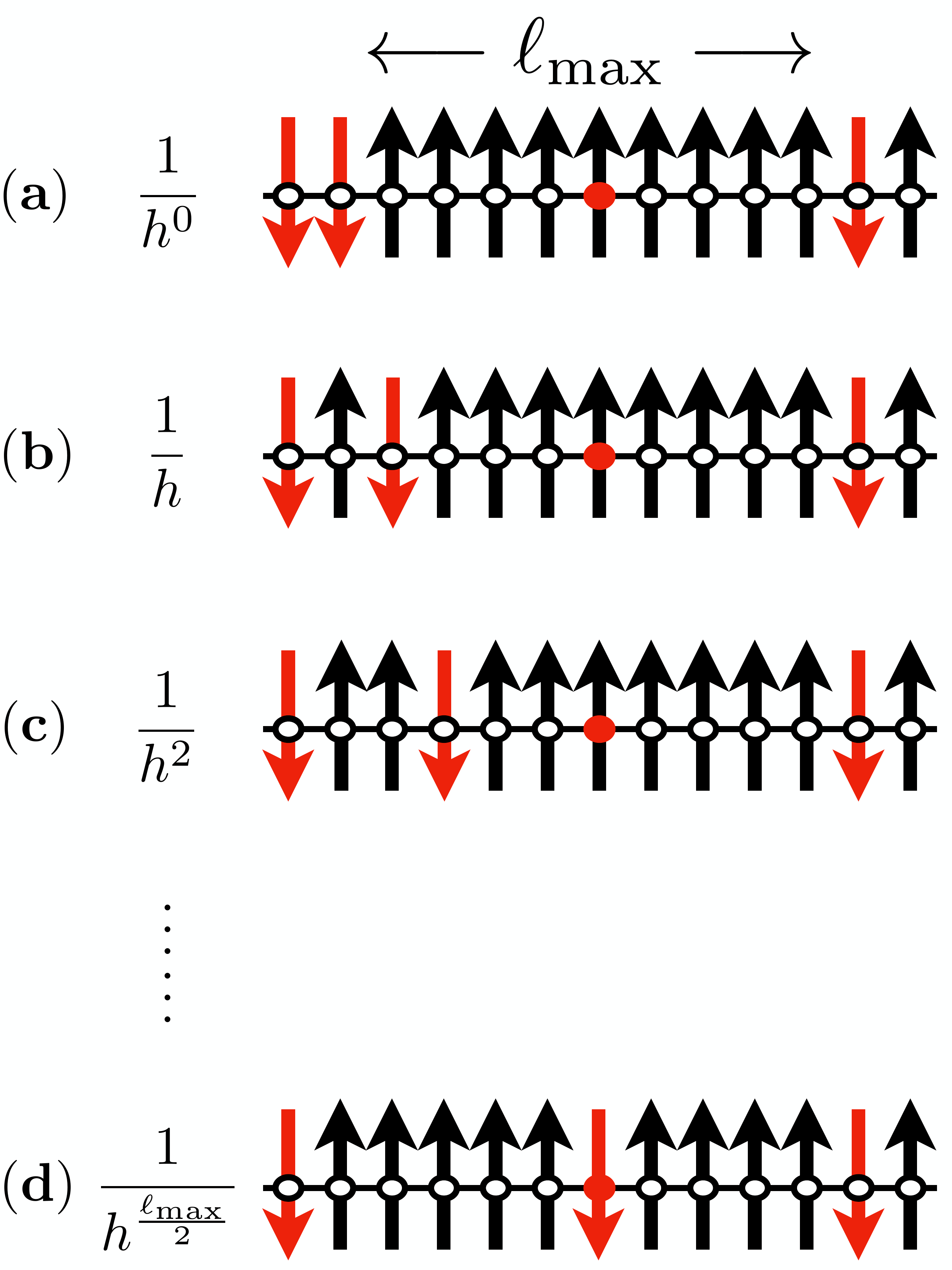}
    \caption{Schematic picture illustrating the high perturbative process necessary to flip the central spin $i_0$ located at the center of the longest polarized region $\ell_\mathrm{max}$. (a) shows the zeroth order perturbative expansion of an MBL eigenstate having a large region ($\ell_\mathrm{max}$) of aligned spins. Spin flip can occur gradually from the boundary of this region, here illustrated from the left side. (b) At first order, the weight is reduced $~\sim h^{-1}$, and (c) at second order $\sim h^{-2}$. Finally (d) shows the $\left(\ell_\mathrm{max}/2\right)$th order (whose weight is strongly reduced $\sim h^{-\ell_\mathrm{max}/2}$) necessary to eventually flip the central spin $i_0$ (red circle).}
    \label{fig:perturbative}
\end{figure}

\textit{MBL regime.---} Following ideas similar to the toy model previously introduced for free fermions, one can get some analytical insights in the MBL case. While the picture of localized single-particle orbitals of Fig.~\ref{fig:orbitals}\,(a) cannot describe the generic interacting XXZ model, we can build perturbative arguments at strong disorder to explain the algebraic decay of the maximally polarized sites.

Deep in the MBL regime, most of the sites display large local magnetizations $|\langle S_i^z\rangle|\simeq 0.5$~\cite{khemani_obtaining_2016,lim_many-body_2016,pietracaprina2018}. Here we argue that the most polarized site $i_0$ belongs to the region which has the largest number $\ell_\mathrm{max}$ of consecutive aligned spins. As schematized on Fig.~\ref{fig:perturbative}, at very large disorder strength $h$, flipping the central spin $i_0$ requires $\ell_\mathrm{max}/2$ spin-flip processes. Performing a perturbative expansion in the spin-flip term, we therefore expect a vanishingly small magnetization,
\begin{equation}
    \frac{1}{2}-|\langle S_{i_0}^z\rangle| \propto h^{-\ell_\mathrm{max}}.
\end{equation}
Estimating $\ell_\mathrm{max}$ is complicated by the presence of interactions. We nevertheless expect $\ell_\mathrm{max}$ to grow with $\log L$, such as in the simpler case of free fermions, but with a non-trivial smaller prefactor, depending on the interaction strength. It follows that $\gamma(h)\propto \log h$ at strong disorder. This is confirmed by the numerical results shown in Fig.~\ref{fig:exponents_MBL}. Similar to Anderson localization (see Fig.~\ref{fig:exponents_anderson}), we also observe here a strong correlation between the exponents $\gamma(h)$ and $\alpha(h)$.

\section{Discussions and conclusions}
\label{sec:summary_conclusions}

\subsection{Summary}

In this paper we have addressed the inverse problem for one-dimensional quantum disordered models without and with interactions, respectively describing Anderson localization, many-body localization (MBL) and its transition toward ergodicity at high energy. Starting from a given many-body state $\Psi_0$, an eigenstate of a local Hamiltonian $\mathcal{H}_0$, we have asked whether $\Psi_0$ could be a good approximate eigenstate of another parent Hamiltonian $\mathcal{H}_\mathrm{P}$ having the exact same local form as $\mathcal{H}_0$. To quantify the goodness of the approximation, we have focused on the energy variance $ \sigma^2_\mathrm{P}=\langle\mathcal{H}_\mathrm{P}^2\rangle_{\Psi_0}-\langle\mathcal{H}_\mathrm{P}\rangle_{\Psi_0}^2$ which can be straightforwardly obtained as the result of a numerical diagonalization of the so-called covariance matrix $\mathsf{C}$, see Eq.~\eqref{eq:cov_mat}. Interestingly, for short-range spin models, such as the XXZ Hamiltonian of Eq.~\eqref{eq:H}, the entries $\mathsf{C}_{ij}$ only depend on the two-point correlators evaluated on the eigenstate $|\Psi_0\rangle$, i.e., $\langle S^z_iS^z_j \rangle_{\Psi_0}-\langle S^z_i\rangle_{\Psi_0}\langle S^z_j\rangle_{\Psi_0}$. Starting from a chain of length $L$, the sole knowledge of these $L(L-1)/2$ correlators is sufficient to build the $(L+1)\times (L+1)$ covariance matrix $\mathsf{C}$, and then access both $\sigma^2_\mathrm{P}$ (its eigenvalues) and $\mathcal{H}_\mathrm{P}$ (its eigenvectors).

Non-interacting Anderson and interacting MBL physics display similar results. Indeed, in both cases we have found emerging parent Hamiltonians whose variances vanish algebraically with system size $\sigma^2_\mathrm{P}(L)\propto L^{-\alpha (h)}$, where the disorder-dependent exponent follows $\alpha(h) \sim \log h$ for large disorder strength $h$. As a result, we observe that quantum localization (interacting or not) leads to the non-uniqueness of parent Hamiltonians. Looking more precisely at the microscopic structure of $\mathcal{H}_\mathrm{P}$, the parent disorder configuration $\{h_i^\mathrm{P}\}$ turns out to be strongly correlated to the initial one $\{h_i^0\}$. Over finite segments $s$ we observe $h^\mathrm{P}_i \simeq h_i^0+\mathcal{C}_s$, where $\mathcal{C}_s$ is a global constant shift, thus forming  a plateau-type structure, with sharp steps between consecutive segments, see Figs.~\ref{fig:hlocal_anderson} and~\ref{fig:parentH_MBL}. Interestingly, the jump positions $i$ correspond to minima in the entanglement entropy and at these very same positions, spins are very close to being perfectly polarized, leading to chain breaks upon increasing system size, see Fig.~\ref{fig:exponents_anderson}.

This chain breaking mechanism has been further investigated through a toy model of localized orbitals for Anderson (Fig.~\ref{fig:orbitals}), and using perturbative arguments for MBL (Fig.~\ref{fig:perturbative}), and in both cases compared to exact numerics. It has been found that the maximally polarized site belongs to the largest region $\ell_\mathrm{max}$ with aligned spins. The logarithmic scaling $\ell_\mathrm{max}\propto\log L$ naturally implies for the decay exponent $\gamma (h)$, which governs how  the maximally polarized site deviates from perfect polarization $\frac{1}{2}-\left|\langle S_i^z\rangle\right|_\mathrm{max}\propto L^{-\gamma(h)}$, to increase with disorder strength as $\gamma(h)\sim \log h$ at large $h$, in the same way as the energy variance exponent $\alpha(h)$. Likewise, the minimal entanglement across this bottleneck decays with the same exponent $\gamma(h)$, see Eq.~\eqref{eq:SEgamma} and Fig.~\ref{fig:exponents_anderson}\,(c).

The situation is completely different for high energy ergodic eigenstates, where, contrary to the (many-body) localized physics, no bottlenecks and chain-breaking events occur. Indeed, we have numerically and analytically found that, disregarding the trivial degrees of freedom due to total energy and total magnetization conservations, there is no parent Hamiltonian with vanishing energy variance for generic thermal states. The smallest non-trivial eigenvalue of the covariance matrix $\mathsf{C}$ goes to the finite value $1/4$ in the thermodynamic limit, a direct consequence of the eigenstate thermalization hypothesis, see Fig.~\ref{fig:scaling_e3_mbleth} and Eq.~\eqref{eq:ETH}.

As a consequence, the ergodic-to-MBL transition can be nicely captured owing to sharply distinct scaling behaviors of the energy variance $\sigma^2_\mathrm{P}(L)$, as clearly visible in Fig.~\ref{fig:scaling_e3_mbleth}.

\subsection{Recap}

Let us now summarize the main findings of our paper, which provide several key insights for quantum localization and ergodicity:

\begin{enumerate}[label=(\roman*)]
    \item Ergodic many-body eigenstates fully and uniquely encode their parent Hamiltonian $\mathcal{H}_0$. In the case of the short-range models considered, only two-point correlations are sufficient to characterize $\mathcal{H}_0$.
    \item Anderson and MBL eigenstates do not uniquely encode $\mathcal{H}_0$. Indeed, as system size is increased, they are better and better approximations of eigenstates of a family of parent Hamiltonians $\mathcal{H}_\mathrm{P}$, which only differ from $\mathcal{H}_0$ by their local disorder configuration.
    \item The formation of entanglement bottlenecks is the key mechanism for building distinct parent Hamiltonians. At such bottlenecks, the entanglement entropy as well as the local spin fluctuations vanish, leading to chain breaking in the thermodynamic limit.
    \item The ergodic-MBL transition can therefore be captured from the physics of the parent Hamiltonian, providing another estimate in addition to the standard ones.
\end{enumerate}

\subsection{Outlook}

Our paper opens several perspectives to address quantum localization problems, in particular, beyond one dimension. For instance the two-dimensional Bose-glass problem at zero temperature~\cite{alvarez_critical_2015,ng_quantum_2015} could be revisited from this point of view. Its interacting localized ground state, accessible by quantum Monte Carlo on much larger system sizes than the ones of exact diagonalization, may be a good representative of an excited state of another Hamiltonian. This might enable the probing ot MBL physics in two dimensions using ground-state techniques.

In order to improve the energy variance of parent Hamiltonians, a promising route would be to enlarge the target space of Hamiltonians, allowing, for instance, randomness in the pairwise couplings. We expect the additional terms to better capture the microscopic structure of the bottlenecks, therefore reducing the energy variance $\sigma_\mathrm{P}^2$.

In the MBL context, we may ask how (if at all) the chain-breaking processes are related to the emergent integrability and the $l$-bit picture~\cite{serbyn_local_2013,huse_phenomenology_2014,imbrie_diagonalization_2016,imbrie_local_2017,rademaker_many-body_2017}. A step toward this goal would consist in making a link with the Kane-Fisher problem~\cite{kane_transport_1992} which also leads to a chain breaking in the presence of isolated impurities in a clean background.

Finally, the chain-breaking mechanism associated with spin freezing is certainly a good prerequisite to further improve strong disorder decimation schemes~\cite{pekker_hilbert-glass_2014,you_entanglement_2016,monthus_strong_2018}, for both MBL at high energy as well as low-temperature Bose-glass physics~\cite{refael_strong_2013,dupuis_glassy_2019}.

\begin{acknowledgments}
    M.D. acknowledges LPT Toulouse for hospitality during the final stage of this work. M.D. was supported by the U.S. Department of Energy, Office of Science, Office of Basic Energy Sciences, Materials Sciences and Engineering Division under Contract No. DE-AC02-05-CH11231 through the Scientific Discovery through Advanced Computing (SciDAC) program (KC23DAC Topological and Correlated Matter via Tensor Networks and Quantum Monte Carlo). This research used the Lawrencium computational cluster resource provided by the IT Division at the Lawrence Berkeley National Laboratory (Supported by the Director, Office of Science, Office of Basic Energy Sciences, of the U.S. Department of Energy under Contract No. DE-AC02-05CH11231). This research also used resources of the National Energy Research Scientific Computing Center (NERSC), a U.S. Department of Energy Office of Science User Facility operated under Contract No. DE-AC02-05CH11231. N.M. and N.L. benefited from the support of the project THERMOLOC ANR-16-CE30-0023-02 of the French National Research Agency (ANR) and by the French Programme Investissements d'Avenir under the program ANR-11-IDEX-0002-02, reference ANR-10-LABX-0037-NEXT. We acknowledge CALMIP (grants 2018-P0677, 2019-P0677) and GENCI (grant 2018-A0030500225) for HPC resources.
\end{acknowledgments}

\bibliography{parent}

\clearpage\newpage
\appendix
\setcounter{figure}{0}
\setcounter{equation}{0}
\renewcommand{\thefigure}{S\arabic{figure}}
\renewcommand{\theequation}{S\arabic{equation}}

\section{Anderson localization}\label{app:distributions_Anderson}

\begin{figure}[h!]
    \centering
    \includegraphics[width=1\columnwidth]{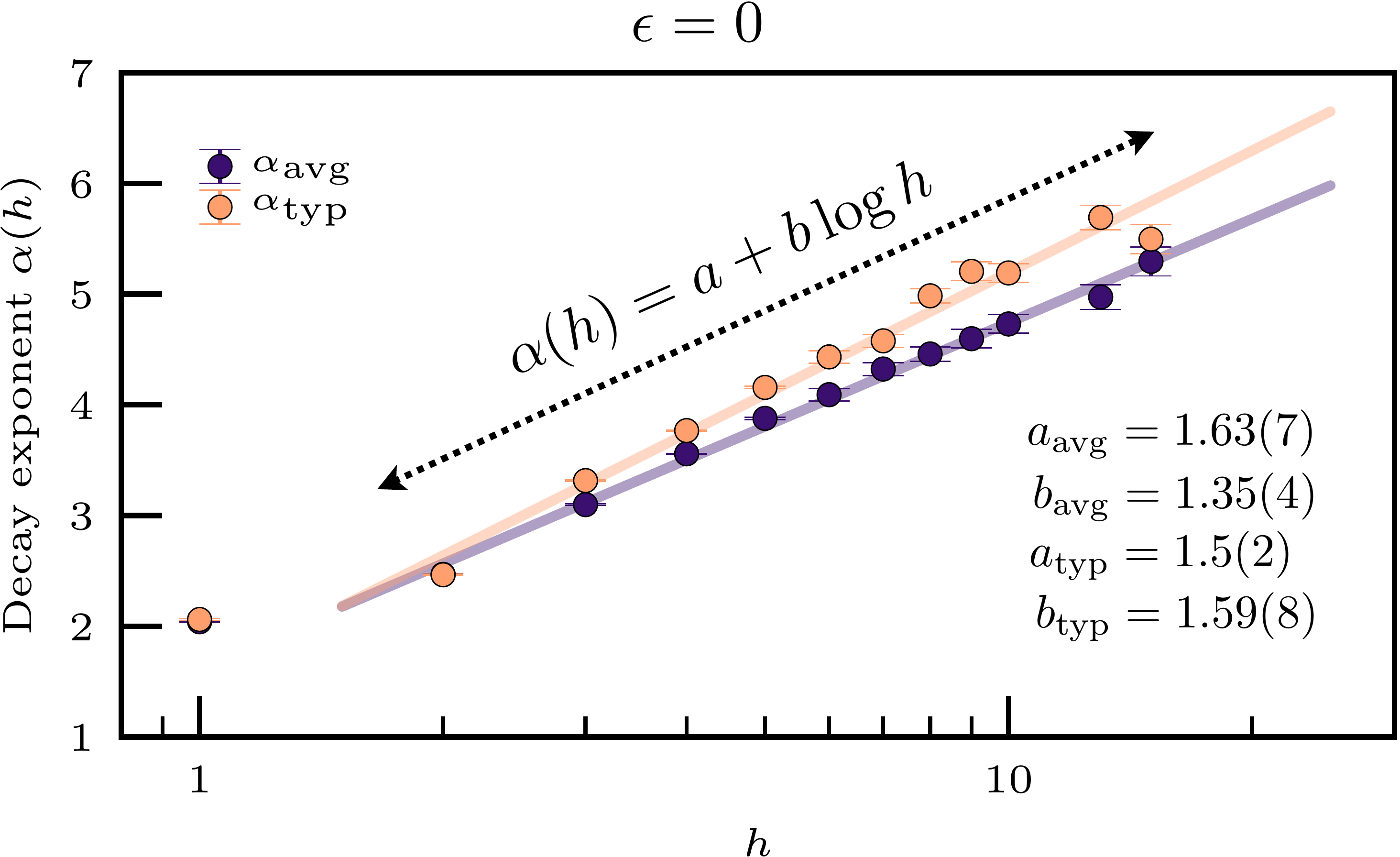}
    \caption{Exact diagonalization results for the ground-state ($\epsilon=0$) of Anderson localization, showing the disorder strength dependence $h$ of the average and typical exponents of the third largest eigenvalue $e_3$ of the covariance matrix $\alpha(h)$, see Eq.~\eqref{eq:e3avgtyp}. A semi-log behavior, with average and typical values very close to one another is observed. Bold lines are fits of the form $a + b\log h$, with the parameters indicated on the plot.}
    \label{fig:exponents_e3_anderson}
\end{figure}

\begin{figure}[h!]
    \centering
    \includegraphics[width=1\columnwidth]{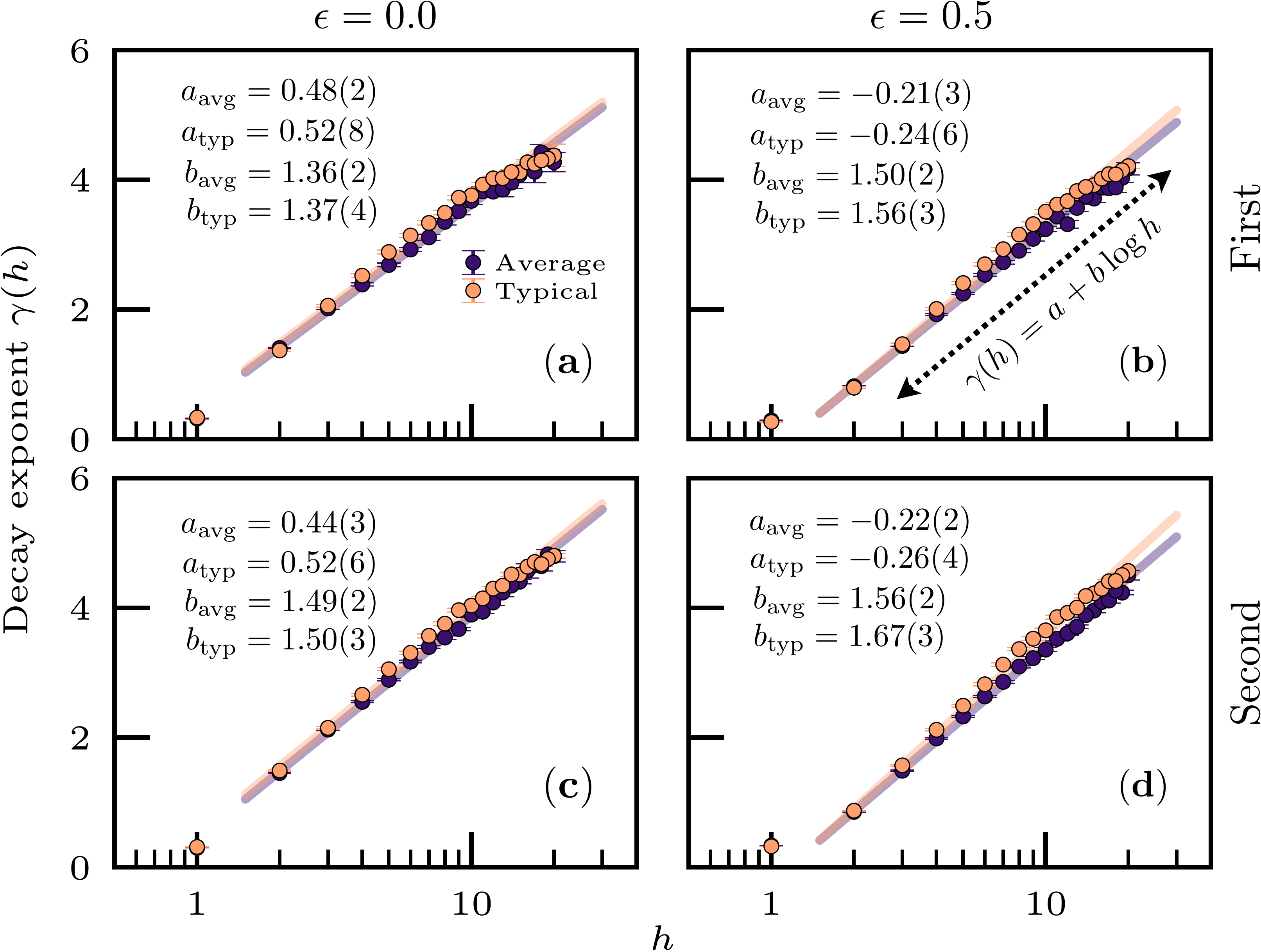}
    \caption{Exact diagonalization results for Anderson localization, showing the $h$-dependence of the average and typical exponents $\gamma(h)$ governing the maximal polarization scaling as in Eq.~\eqref{eq:mzavgtyp}. The upper panels show the largest polarization and the bottom panels the second largest. Left column: Ground state $\epsilon=0$. Right column: High-energy eigenstate $\epsilon=0.5$. One sees in all cases a semi-log behavior, with average and typical values very close to one another, in contrast with MBL. Bold lines are fits of the form $a + b\log h$, with the parameters indicated on the plot.}
    \label{fig:scaling_szmax_anderson}
\end{figure}

Here, we show how average and typical values of the different quantities considered in this paper behave in a very similar way for Anderson localized eigenstates. In particular, we show in Figs.~\ref{fig:exponents_e3_anderson} and~\ref{fig:scaling_szmax_anderson} respectively that the decay exponents with system size of the third smallest eigenvalue $e_3$ of the covariance matrix~\eqref{eq:e3avgtyp}, and of the maximal polarization~\eqref{eq:mzavgtyp} have identical scalings for typical and average values, at variance with high-energy MBL results, see discussions in Sec.~\ref{sec:MBL}. In addition, the bottom panels of Fig.~\ref{fig:scaling_szmax_anderson} show the exponent $\gamma(h)$ for the second most polarized site, which is similar to the most polarized shown in the top panels of Fig.~\ref{fig:scaling_szmax_anderson}.

\section{Many-body localization}\label{app:MBL}

\subsection{Other eigenvalues of the covariance matrix $e_{i \geq 3}$} \label{app:alpha_i}

\begin{table}[b!]
    \begin{center}
        \begin{tabular}{l|c|c|c|c}
            $i$&3&4&5&6\\
            \hline
            $\alpha_\text{avg}$&4.67(3)&4.77(3)&4.90(3)&5.07(2)\\
            $\alpha_\text{typ}$&5.4(1)&5.1(1)&5.2(1)&5.5(1)\\
            \hline
            \hline
        \end{tabular}
        \caption{\label{tab:alpha_i}Decay exponent $\alpha_i$ of the $i$th eigenvalue of the covariance matrix, estimated from the average and typical values at $h=15$ shown in Fig.~\ref{fig:decay_ei_MBL}.}
    \end{center}
\end{table}

\begin{figure}[h!]
    \centering
    \includegraphics[width=\columnwidth]{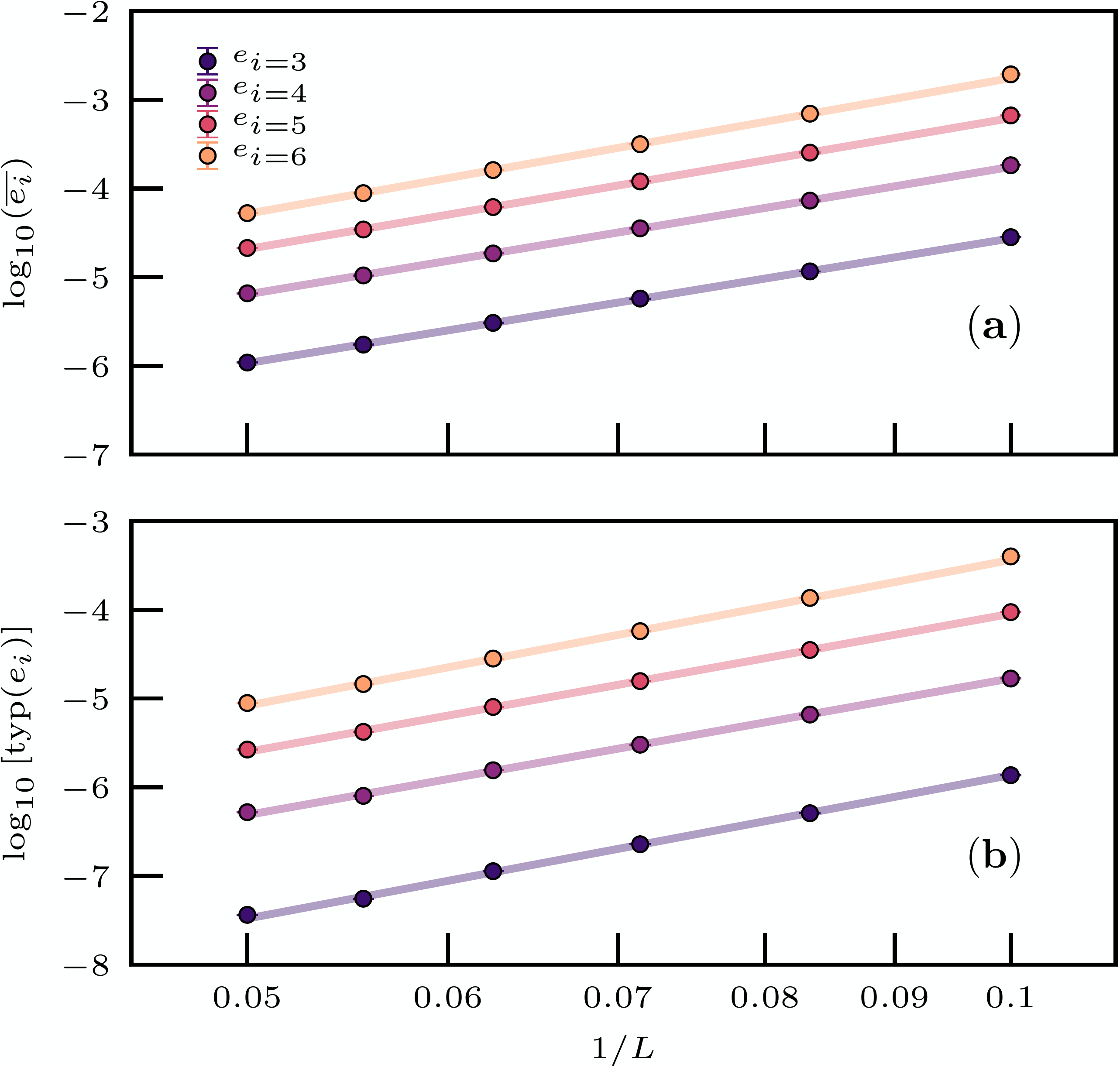}
    \caption{Scaling of the smallest eigenvalues $e_i$ of the covariance matrix with the inverse system size $1/L$, in the MBL phase at disorder $h=15$. (a) Scaling of the average values. (b) Scaling of the typical values. The data are obtained from at least $20\,000$ samples, using one to ten eigenstate(s) per sample.}
    \label{fig:decay_ei_MBL}
\end{figure}

In Fig.~\ref{fig:decay_ei_MBL} we look at the first few non-zero eigenvalues of the covariance matrix $e_{i\geq 3}$ at high energy in the MBL phase. As in the case of the ground state~\cite{dupont2019}, we observe that they all vanish algebraically with the system size. This further confirms that, as argued in Sec.~\ref{subsec:consequences}, the chain breaking mechanism happens not only at one site, but at a set of sites $\{i_0\}$ in the thermodynamic limit. The average decay exponents given in Table~\ref{tab:alpha_i} are seen to slightly increase with index $i$, while the typical decay exponents are almost independent of the index $i$.

\subsection{Strong disorder distributions}\label{app:distributions_MBL}

\begin{figure}[b!]
    \centering
    \includegraphics[width=\columnwidth]{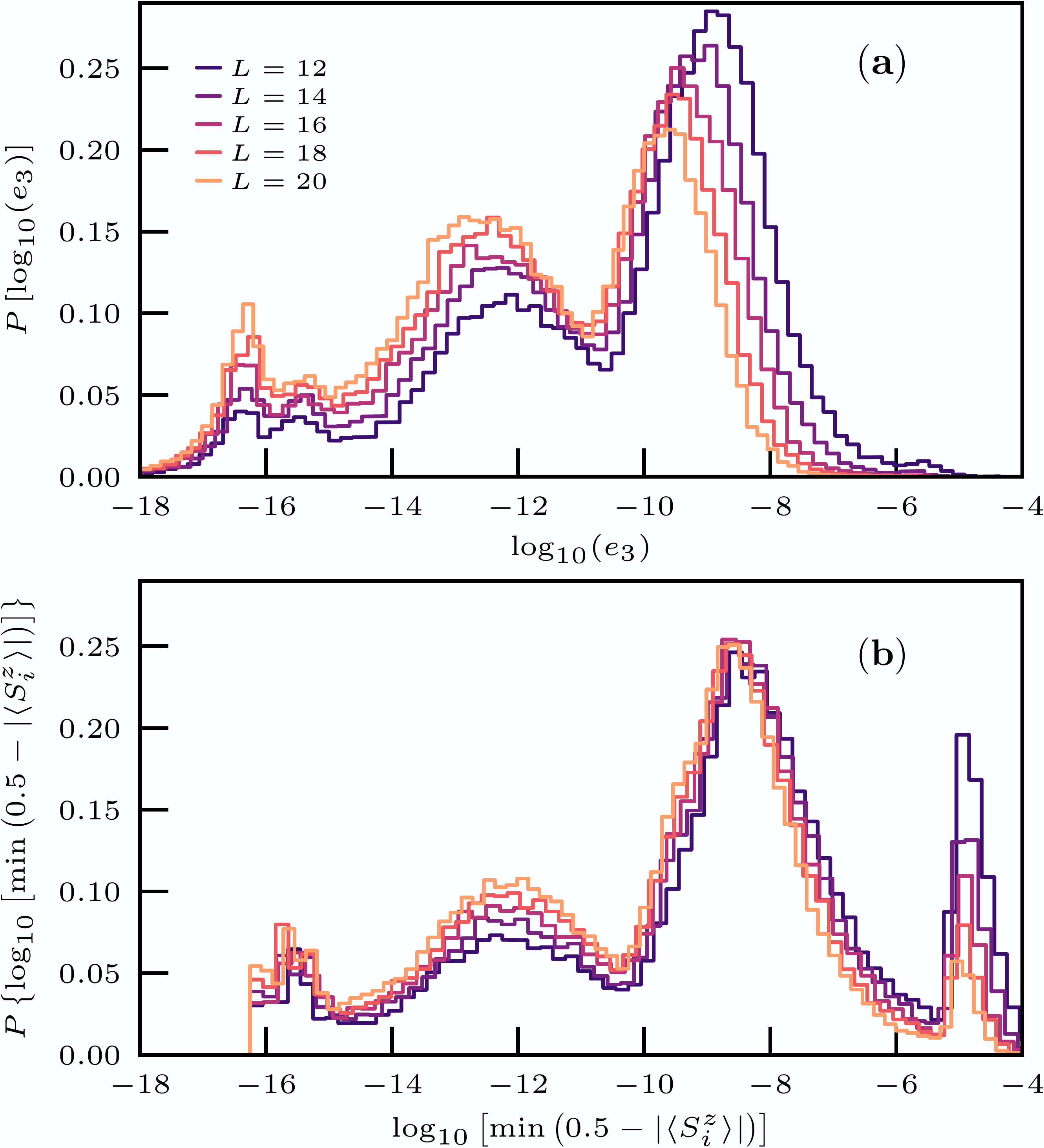}
    \caption{Distributions deep in the MBL phase, at disorder $h=100$. (a) Distribution of the variance $e_3$. (b) Distribution of the maximal polarization.}
    \label{fig:distributions_MBL}
\end{figure}

To understand the relationship between the different exponents at strong disorder in the MBL phase, given in Fig.~\ref{fig:exponents_MBL}, it is useful to look at the distribution of the corresponding observables. In that respect, Fig.~\ref{fig:distributions_MBL} shows that both the energy variance $e_3$ and the maximal polarization feature several peaks, have a large negative skewness, and are not self-averaging. Non-zero skewness entails that average and typical values are different, and the non self-averaging character indicates that this difference will persist even in the thermodynamic limit. For both quantities, as system size is increased, weight is transferred from large to small values.
In the case of the variance, see Fig.~\ref{fig:distributions_MBL}\,(a), this weight transfer comes with a very visible shift of the whole distribution toward smaller values. In the case of the maximal polarization shown in Fig.~\ref{fig:distributions_MBL}\,(b), such a shift is also visible but is far less important. We can thus expect the average maximal polarization exponent -- mainly influenced by a shift of the distribution, which here is tiny --- to be significantly smaller than the typical exponent --- mainly influenced by weight transfer. This is confirmed by the numerical computation of the exponents plotted in the inset of Fig.~\ref{fig:exponents_MBL}, which yields $\gamma_\text{typ}(h)\simeq 2\gamma_\text{avg}(h)$. By contrast, in the case of the variance both the weight transfer and shift are important, and we accordingly observe that $\alpha_\text{typ}(h)\simeq\alpha_\text{avg}(h)$.

\clearpage\newpage

\end{document}